\documentclass[nofootinbib,a4paper,aps,prA,reprint,twocolumn,preprintnumbers,amsmath,amssymb,10pt, superscriptaddress]{revtex4-1}
\usepackage{amsmath}
\usepackage{verbatim}
\usepackage{graphicx}
\usepackage{color}
\usepackage{tikz}
\usepackage[colorlinks=true,citecolor=blue,linkcolor=blue,urlcolor=blue]{hyperref}
\usepackage{braket, enumitem,}
\usepackage[normalem]{ulem}
\usepackage{upgreek}

\begin{document}

\title{Spectral properties of critical 1+1D Abelian-Higgs model}

\author{Titas Chanda} 
\email{titas.hri@gmail.com}
\affiliation{Department of Physics, Indian Institute of Technology Indore, Khandwa Road, Simrol, Indore 453552, India}
\affiliation{The Abdus Salam International Centre for Theoretical Physics (ICTP), Strada Costiera 11, 34151 Trieste,
Italy}

\author{Marcello Dalmonte}
 \affiliation{The Abdus Salam International Centre for Theoretical Physics (ICTP), Strada Costiera 11, 34151 Trieste,
Italy}
\affiliation{International School for Advanced Studies (SISSA), via Bonomea 265, 34136 Trieste, Italy}

\author{Maciej Lewenstein}
\affiliation{ICFO-Institut de Ci\`encies Fot\`oniques, The Barcelona Institute of Science and Technology, Av. Carl Friedrich
Gauss 3, 08860 Barcelona, Spain}
\affiliation{ICREA, Passeig Lluis Companys 23, 08010 Barcelona, Spain}

\author{Jakub Zakrzewski}
\affiliation{Instytut Fizyki Teoretycznej, Uniwersytet Jagiello{\'n}ski, \L{}ojasiewicza 11, 30-348 Krak{\'o}w, Poland}
\affiliation{Mark Kac Complex Systems Research Center, Jagiellonian University in Krakow, \L{}ojasiewicza 11, 30-348 Krak\'ow,
 Poland.}

\author{Luca Tagliacozzo}
\email{luca.tagliacozzo@iff.csic.es}
\affiliation{Instituto de Física Fundamental IFF-CSIC, Calle Serrano 113b, Madrid 28006, Spain}

\begin{abstract}
The presence of  gauge symmetry in 1+1D is known to be redundant, since it does not imply the existence of dynamical gauge bosons. As a consequence, in the continuum, the Abelian-Higgs model, the theory of bosonic matter interacting with photons, just possesses a single phase, as the higher dimensional Higgs and Coulomb phases are connected via non-perturbative effects. However, recent research published in [Phys. Rev. Lett. \textbf{128}, 090601 (2022)] has revealed an unexpected phase transition when the system is discretized on the lattice. This transition is described by a conformal field theory with a central charge of $c=3/2$. In this paper, we aim to characterize the two components of this $c=3/2$ theory -- namely the free Majorana fermionic and bosonic parts -- through equilibrium and out-of-equilibrium spectral analyses. 
\end{abstract}

\maketitle

\section{Introduction}

A physical theory is most often characterized by fixing its symmetries, the dimensions of space-time it operates on, and its field content~\cite{Coleman1985, DiFrancesco1997,Blumenhagen2009,Shankar2017,Fradkin2013}. 
This concept forms the basis of the renormalization group construction and our efforts to categorize its fixed points~\cite{wilson71, wilson71b, wilson75, Shankar94, gellmann_low}. For instance, in a four-dimensional space-time, in a system featuring $U(1)$ gauge-invariance and fermionic matter, we expect the physics to be described by quantum electrodynamics, where the fermions correspond to electrons and the gauge bosons correspond to photons~\cite{peskin_book, Schwartz2013}.

Here we analyze a scenario that shows that this simple picture does not necessarily hold. The lattice discretization of the Abelian-Higgs model~\cite{anderson_gauge_63, englert_broken_64, higgs_broken_64, Guralnik_conservation_64, peskin_book, Schwartz2013} (see Refs.~\cite{endres_higgs_2012, kasamatsu2013, pekker_amplitude_2015, kuno2017, gonzalez-cuadra2017, zhang18, unmuth-yockey_universal_2018,park2019, Meurice21} for recent discussions on realizing Abelian-Higgs model in quantum simulators), defined on one spatial and one time dimensions (1+1D), shares the same dimensionality and field content with the same in the continuum. However, as shown recently in Ref.~\cite{chanda2021}, strong correlations induce an emergent physical theory on the lattice that can be described, at low energies, as one massless relativistic fermion together with a massless relativistic boson.

The Abelian-Higgs model is among the simplest $U(1)$ gauge theory in 1+1D. It contains bosonic matter (the Higgs part of the model) and constitutes a well-known textbook example of gauge theory \cite{tong, peskin_book, Schwartz2013}. The model is described by the action (with the metric convention $g_{\mu \nu} = \text{diag}(-1, 1)$):
\begin{align}
S = \int d^2x  \Big( &-\frac{1}{2} F^{01}F_{01} - D^{\nu}\phi \left(D_{\nu}\phi\right)^*  \nonumber \\
&- m^2 |\phi|^2
-\frac{\lambda}{2}|\phi|^4+\frac{\theta}{2\pi} q F^{01}\Big),
\end{align}
where
 $D_{\nu}=\partial_{\nu}+iq A_{\nu} $ is the covariant derivative, with $q$ being the charge of the scalar matter field described by the complex-valued field $\phi$. $m$ and $\lambda$ are the bare mass and self-interaction of the scalar field, respectively, and $\theta$ is a periodic parameter, $-\pi\le \theta \le \pi$, that controls the background field. 

Despite the similarity of the above action with higher-dimensional Abelian-Higgs models, the physics in 1+1D is very different~\cite{Coleman1985,komargodski_comments_2019, tong}. In higher dimensions, the model displays two well-known phases, a Coulomb phase described by massless photons interacting with the matter field, and a Higgs phase where the photons are screened and acquire mass, and the Higgs field condenses~\cite{fradkin_phase_1979, Callaway_phase_1982}.  In 1+1D, for positive $m^2$ the scalar matter is confined. Differently from the higher-dimensional models, instanton effects lead to confinement also for large and negative $m^2$, destroying the expected screening  caused by the condensation of the Higgs field~\cite{Coleman1985,komargodski_comments_2019, tong}.
The only exception to this simple scenario involving the occurrence of a single gapped phase in the continuum is at $\theta =\pi$ (see e.g., \cite{tong, Gattringer2018, Goschl2018, Sulejmanpasic2020, sulejmanpasic21}), where there is a line of first-order transition that culminates into a second-order critical point in the Ising universality class (akin to what happens in the Schwinger model~\cite{Coleman1975Schwinger, Coleman1976}). 

In the recent works~\cite{chanda20, chanda2021}, the same model has been discretized on the lattice in the Hamiltonian formalism. Having $L$ sites (with lattice spacing $a$) and $L-1$ bonds the corresponding Hamiltonian, at $\theta=0$, reads 
\begin{align}
\hat{H} = \sum_j \Big[ \hat{L}^2_j &+ 2 x \ \hat{\Pi}^{\dagger}_j \hat{\Pi}_j + (4 x - \frac{2 \mu^2}{q^2}) \hat{\phi}^{\dagger}_j\hat{\phi}_j    \nonumber \\
&+ \frac{\lambda}{q^2} (\hat{\phi}_j^{\dagger})^2 \hat{\phi}_j^2 
- 2 x (\hat{\phi}^{\dagger}_{j+1} \hat{U}_j \hat{\phi}_j + \text{h.c.})\Big],
\label{eq:Hamil}
\end{align}
 with  $x = 1/a^2 q^2$ and $\mu^2 = -m^2$. The 
 matter field operators, $\{\hat{\phi}_j,  \hat{\phi}_j^{\dagger},
\hat{\Pi}_j,  \hat{\Pi}_j^{\dagger}\}$,  act on the Hilbert space at site $j$, while the gauge-field operators, $\{\hat{L}_j,  \hat{U}_j, \hat{U}^{\dagger}_j\}$, act on  Hilbert space defined on the bond linking sites $j$ and $j+1$.
The operators fulfil the standard canonical commutation relations:
\begin{align}
    \ [\hat{\phi}_j, \hat{\Pi}_k] = [\hat{\phi}^{\dagger}_j, \hat{\Pi}^{\dagger}_k] = i \delta_{jk};  \nonumber \\
    [\hat{L}_j, \hat{U}_j] = -\hat{U}_j, \ \ [\hat{L}_j, \hat{U}^{\dagger}_j] = \hat{U}^{\dagger}_j.
\end{align}
The usual continuum limit is taken as $x \rightarrow \infty$.

Rather than considering the usual continuum limit,
$x=2$ was fixed in \cite{chanda2021}, and the phase diagram as a function of $\lambda$ and $\mu$ was characterized numerically with matrix product state (MPS) techniques~\cite{schollwock_aop_2011,orus_aop_2014} (see Refs. \cite{banuls_jhep_2013, buyens_prl_2014, kuhn_pra_2014, banuls_prd_2015, buyens_prd_2016, pichler_prx_2016, banuls_prl_2017, buyens_prx_2016, Magnifico2020, Banuls2020, Aidelsburger2021} for recent applications of MPS techniques in models of lattice gauge theory). On the lattice, the model presents a rich phase diagram. Even at $\theta=0$, there is first-order phase transition between two distinct regimes, a confined regime characterized by the suppression of the kinetic term of the Hamiltonian $(\hat{\phi}^{\dagger}_{j+1} \hat{U}_j \hat{\phi}_j + \text{h.c.})$ and low entanglement entropy in the ground state, and a gapped ``Higgs'' regime, characterized by a larger (but finite) kinetic energy and entanglement entropy.
The first-order line ends in a second-order critical point. 
This critical point is described by a conformal field theory (CFT) with central charge of $c=3/2$. 
This implies that the interplay of lattice effects and non-perturbative physics can give rise to new emerging phenomena, whose field theory description is completely different from the original continuum field theory discretized to obtain the lattice model.

As a result, in the continuum, we have two completely different models emerging from the same microscopic model. A gapped model of confined bosonic charges obtained by taking $x \to \infty$ in any region of the phase diagram on the lattice model whose physics is well known and described in textbooks~\cite{peskin_book,tong,Schwartz2013}. At the critical point on the lattice, however, we can take a different continuum limit at fixed $x$ and obtain a gapless conformal field theory with $c=3/2$ that originates from the same microscopic model -- relativistic bosons interacting with a $U(1)$ gauge field. This continuum theory is less understood. 

In this work, we proceed to better characterize this 
lattice microscopic model. We start solving some of the puzzles presented in Ref.~\cite{chanda2021}.
In particular, we show that all Renyi entropies, differently from what was claimed originally, scale as expected from the conformal field theory predictions, once the sub-leading corrections are taken into account.

We then extract the sound velocity and the dispersion relation of the system via real-time dynamics, and combine those with an accurate finite size-scaling analysis of the Hamiltonian spectrum. From those, we identify the presence of a charge gap, and identify the portion of the low energy spectrum that is responsible for critical behavior.  Since the model is gauge invariant, we can target different background charge sectors. In the zero background charge sector, we only find the Majorana fermionic or Ising part of the spectrum.  
On the other hand, using known relation about the fluctuations of the charges, we also identify the value of the Luttinger liquid parameter of the bosonic sector of the theory. 

The resulting picture is thus much clearer than before but still incomplete. As far as we know, the critical point of the discretized Abelian Higgs model in 1+1D provides the first example of a relativistic CFT where  a discrepancy between the scaling of entanglement and the low-energy spectrum is observed. 

We thus extend our analysis to the other sectors of background charges, where we find many low-energy excitations that should encode the bosonic part of the spectrum.  However, with our current tools, we have limited precision in extracting the spectra in the other background charge sectors to properly identify all of them. Thus  we are still not able to completely identify the full low-energy spectrum, something we plan to do in the future.

\section{A Brief Recap: The system}
\label{sec:model}

The Abelian-Higgs model in 1+1D is described by the Hamiltonian \eqref{eq:Hamil}. We can then define creation and annihilation operators for particles `$a$'  and anti-particles `$b$'  as $\hat{a}_j$ and $\hat{b}_j$ fulfilling
$[\hat{a}_j, \hat{a}^{\dagger}_k] = [\hat{b}_j, \hat{b}^{\dagger}_k] = \delta_{jk}$ The operators are defined as
\begin{eqnarray*}
\hat{\phi}_j = \frac{1}{\sqrt{2}} \left(\hat{a}_j + \hat{b}_j^{\dagger}\right), \ \hat{\Pi}_j = \frac{i}{\sqrt{2}} \left(\hat{a}^{\dagger}_j - \hat{b}_j\right), \\
\hat{\phi}^{\dagger}_j = \frac{1}{\sqrt{2}} \left(\hat{a}^{\dagger}_j + \hat{b}_j\right), \ \hat{\Pi}^{\dagger}_j = \frac{i}{\sqrt{2}} \left(\hat{b}^{\dagger}_j - \hat{a}_j\right),
\end{eqnarray*} as discussed in e.g., \cite{chanda20}.

\subsection{Gauge invariance and Gauss law generators}

In the absence of external charges, the local $U(1)$ symmetry implies that all physical states $|\Psi\rangle$ satisfy Gauss law  $\hat{G}_j |\Psi\rangle =0,\ \forall j$, where the generators are~\cite{chanda20}:
\begin{equation}
\hat{G}_j = \hat{L}_j - \hat{L}_{j-1} -{\hat{Q}_j},    
\end{equation}
with
$\hat{Q}_j = \hat{a}_j^{\dagger} \hat{a}_j - \hat{b}_j^{\dagger} \hat{b}_j$ describing the dynamical charge.
Using Gauss law, we can integrate-out the gauge fields in a chain with open boundary conditions 
by the non-local transformations:
\begin{align}
    \left( \prod_{l \leq j} \hat{U}_j \right) \hat{\phi}_j \rightarrow \hat{\phi}_j, \ 
    \left( \prod_{l \leq j} \hat{U}^{\dagger}_j \right) \hat{\Pi}_j \rightarrow \hat{\Pi}_j, \
    \hat{L}_j = \sum_{l \leq j} \hat{Q}_l.
    \label{eq:nonlocal_transformation}
\end{align}
This introduces a long-range potential for the matter fields \cite{schwinger_pr_1951} as follows:
\begin{align}
\hat{H} = \sum_j \Big[ \left(\sum_{l\leq j} \hat{Q}_l\right)^2 + 2 x \ \hat{\Pi}^{\dagger}_j \hat{\Pi}_j + (4 x - \frac{2 \mu^2}{q^2}) \hat{\phi}^{\dagger}_j\hat{\phi}_j   \nonumber \\
+ \frac{\lambda}{q^2} (\hat{\phi}_j^{\dagger})^2 \hat{\phi}_j^2  - 2 x (\hat{\phi}^{\dagger}_{j+1} \hat{\phi}_j + \text{h.c.})\Big].
\label{eq:Hamil_pureM}
\end{align}

\begin{figure}
     \centering
     \includegraphics[width=\linewidth]{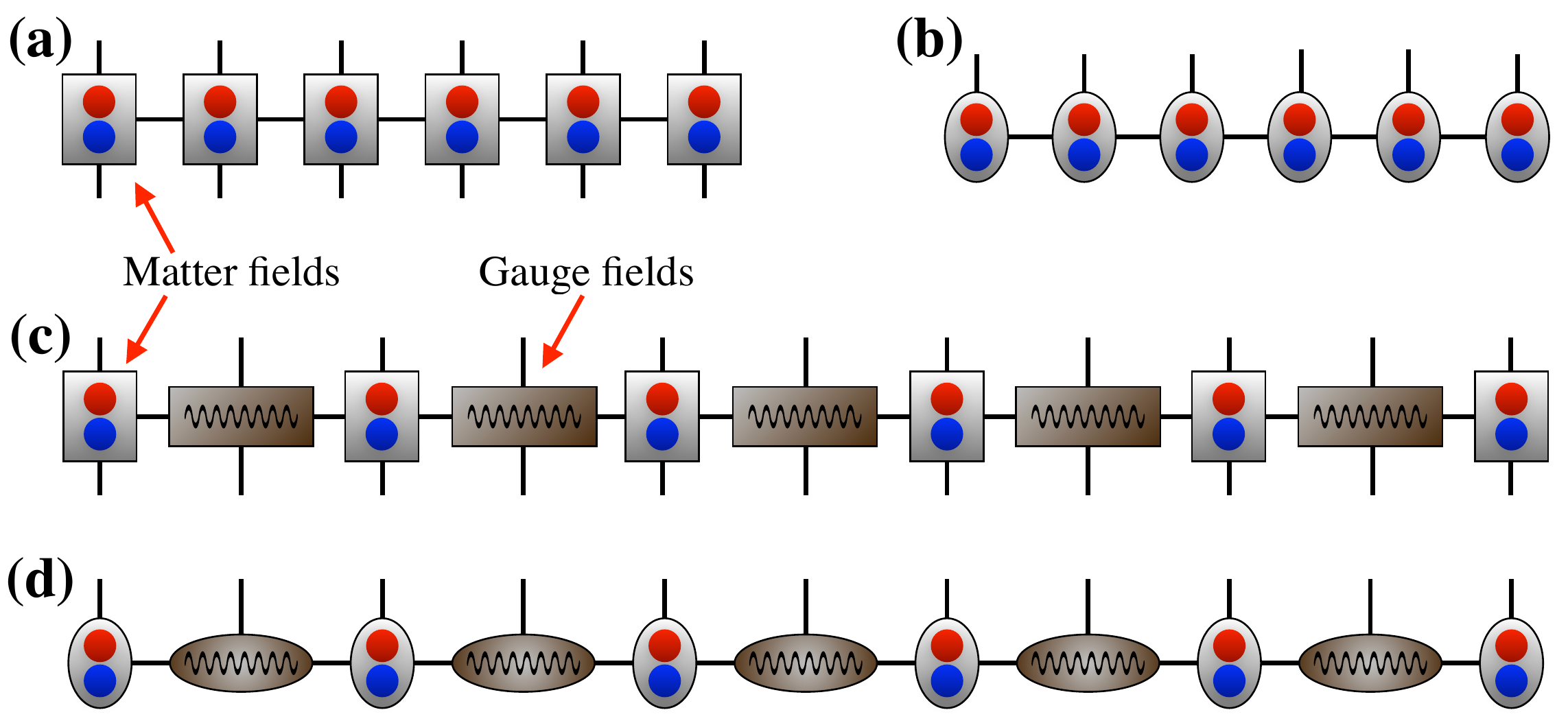}
     \caption{(Color online.) Diagrammatic depictions of the MPO (panels \textbf{(a)} and  \textbf{(c)} and MPS (panels \textbf{(b)} and  \textbf{(d)} representations used in this paper. Panels \textbf{(a)} and \textbf{(b)} correspond to the long-range matter-only description of Eq.~\eqref{eq:Hamil_pureM}, while \textbf{(c)} and \textbf{(d)} presents the same for the matter-gauge picture of Eq.~\eqref{eq:Hamil}.
     }
     \label{fig:mps}
 \end{figure}

In the rest of the paper, unless explicitly stated otherwise, we shall consider Eq.~\eqref{eq:Hamil_pureM} as the system Hamiltonian\footnote{It is to be noted that since the mapping in Eq.~\eqref{eq:nonlocal_transformation} is exact, the long-range Hamiltonian \eqref{eq:Hamil_pureM} describes the same physics as that of the short-range matter-gauge Hamiltonian \eqref{eq:Hamil}. We verify this numerically in Appendix~\ref{app:comparison}.}  that respects global $U(1)$ symmetry corresponding to the conservation of the total dynamical charge $\hat{Q} = \sum_j \hat{Q}_j$. 
We use the density matrix renormalization group (DMRG) algorithm \cite{schollwock_aop_2011,orus_aop_2014,white_prl_1992,white_prb_1993,white_prb_2005,schollwock_rmp_2005} in the framework of the matrix product states (MPS)
 to find the ground state or low-lying excited states of the  Hamiltonian~\eqref{eq:Hamil_pureM}. 
 The diagrammatic depictions of the matrix product operator (MPO) for the Hamiltonian~\eqref{eq:Hamil_pureM} and the corresponding MPS are given in Fig.~\ref{fig:mps}\textbf{(a)} and \textbf{(b)} respectively.
 In Fig.~\ref{fig:mps}\textbf{(c)} and \textbf{(d)} we also present the same for the short-range matter-gauge description of Eq.~\eqref{eq:Hamil}, where we also impose global $U(1)$ symmetry to conserve of the total dynamical charge $\hat{Q}$.
 For time-evolution, we use MPS based time-dependent variational principle (TDVP) \cite{haegeman_prl_2011, haegeman_prb_2016, koffel_prl_2012, paeckel_aop_2019} concentraing only on the long-range description.

\subsection{The phase diagram} 

\begin{figure}[tbh]
\centering
\includegraphics[width=\linewidth]{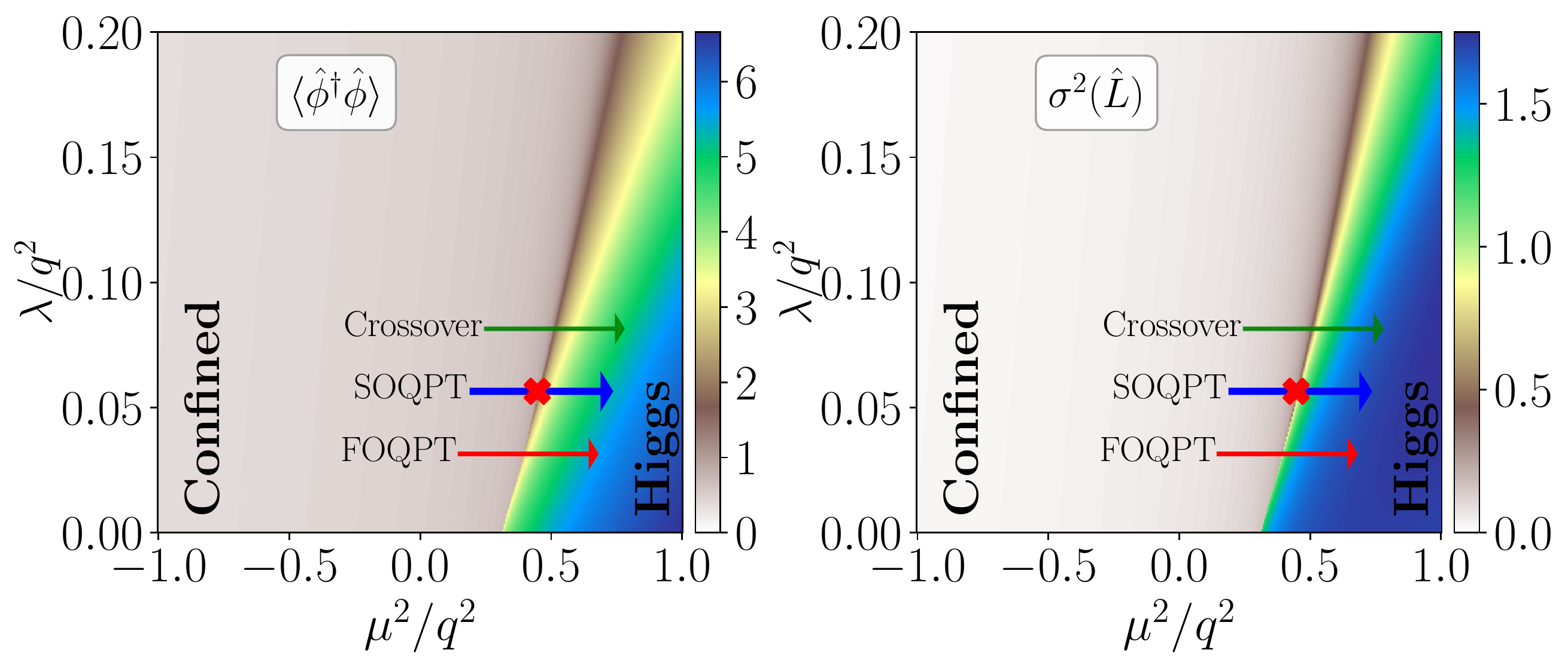}
\caption{Phase diagram of the Abelian-Higgs model in 1+1D \eqref{eq:Hamil_pureM} in the $(\mu^2/q^2, \lambda/q^2)$-plane for a system of size $L=60$. The left panel shows the behavior of the expectation value $\braket{\hat{\phi}^{\dagger} \hat{\phi}}$ while the right panel is for the variance $\sigma^2(\hat{L})$. Both the quantities are averaged over lattice sites/bonds.}
\label{fig:phase_diagram}
\end{figure}

The phase diagram of the Abelian-Higgs model in 1+1D on a discrete lattice has been studied recently in~\cite{chanda2021}. Here we present a different characterization leading to the same phase-diagram. In continuum, the phase diagram of the  Abelian-Higgs model is somewhat uninteresting as there exists only one gapped confined phase without any phase transition in absence of background field. However, the lattice discretization unveils a rich landscape of phase transitions between the confined and the Higgs phase. At weak coupling (i.e., at small $\lambda/q^2$) these two phases are separated by a line of first order quantum phase transition (FOQPT) that ends at a critical second order quantum phase transition (SOQPT). Above this SOQPT point these two phases are smoothly connected by a crossover. In Fig.~\ref{fig:phase_diagram}, we depict the phase diagram of system in the $(\mu^2/q^2, \lambda/q^2)$-plane.

In the Higgs phase, since the gauge field becomes massive, its quantum fluctuations as measured by the variance $\sigma^2(\hat{L}_j) = \braket{\hat{L}^2_j} - \braket{\hat{L}_{j}}^2$ becomes non-zero, while it remains vanishingly small in the confined region. Oppositely, since the matter field attains finite expectation value in the Higgs phase, $\braket{\hat{\phi}_j^{\dagger} \hat{\phi}_j}$ becomes large as seen in Fig.~\ref{fig:phase_diagram}.

The SOQPT point has been located precisely to be $(\lambda_c/q^2 = 0.0565(1), \ \mu_c^2/q^2 = 0.447(1))$ in \cite{chanda2021} using  predictions about  scaling of entanglement entropy for a CFT. There it was also shown that the entanglement entropy and the kinetic term in the Hamiltonian could be used to illustrate the phase diagram as we have done here with the fluctuation of the electric field and the onsite occupation of the bosons. 

The analysis of Ref.~\cite{chanda2021} using the well-established machinery of CFT predicted the existence of a direct sum of two critical theories, free Majorana fermion and free boson, at the critical point. In this work, we complement the previous work with a proper characterization of these two gapless modes.

\subsection{Sound velocity at criticality}
\label{sec:sound_velocity}

In order to verify that the critical point $(\mu_c^2/q^2 = 0.447, \lambda_c/q^2 = 0.0565)$ displays Lorentz invariance in its low-energy description,  we calculate the sound velocity $v_s$ at the critical point from the scaling of ground state energy -- recapitulating the result presented in the supplementary material of Ref.~\cite{chanda2021}. For a Lorentz invariant system exhibiting a linear dispersion at low energies, the ground state energy $E_0$, under open boundary conditions, scales with the system size $L$ according to \cite{Blote_1986, Affleck_1986, Affleck_1989}:
\begin{equation}
E_0(L) = \epsilon_0^{\infty} L + \epsilon_1^{\infty} - \frac{\pi c v_s}{24 L},
\label{eq:gs_scaling}
\end{equation}
where $\epsilon_0^{\infty}$ is the ground state energy density in the bulk and $\epsilon_1^{\infty}$  is the surface free energy in the thermodynamic limit, $c$ is the central charge of the corresponding CFT, and $v_s$ is the sound velocity.  It should be noted that $v_s$ vanishes for Lorentz non-invariant critical points characterized by a quadratic dispersion.

\begin{figure}
    \centering
    \includegraphics[width=0.8\linewidth]{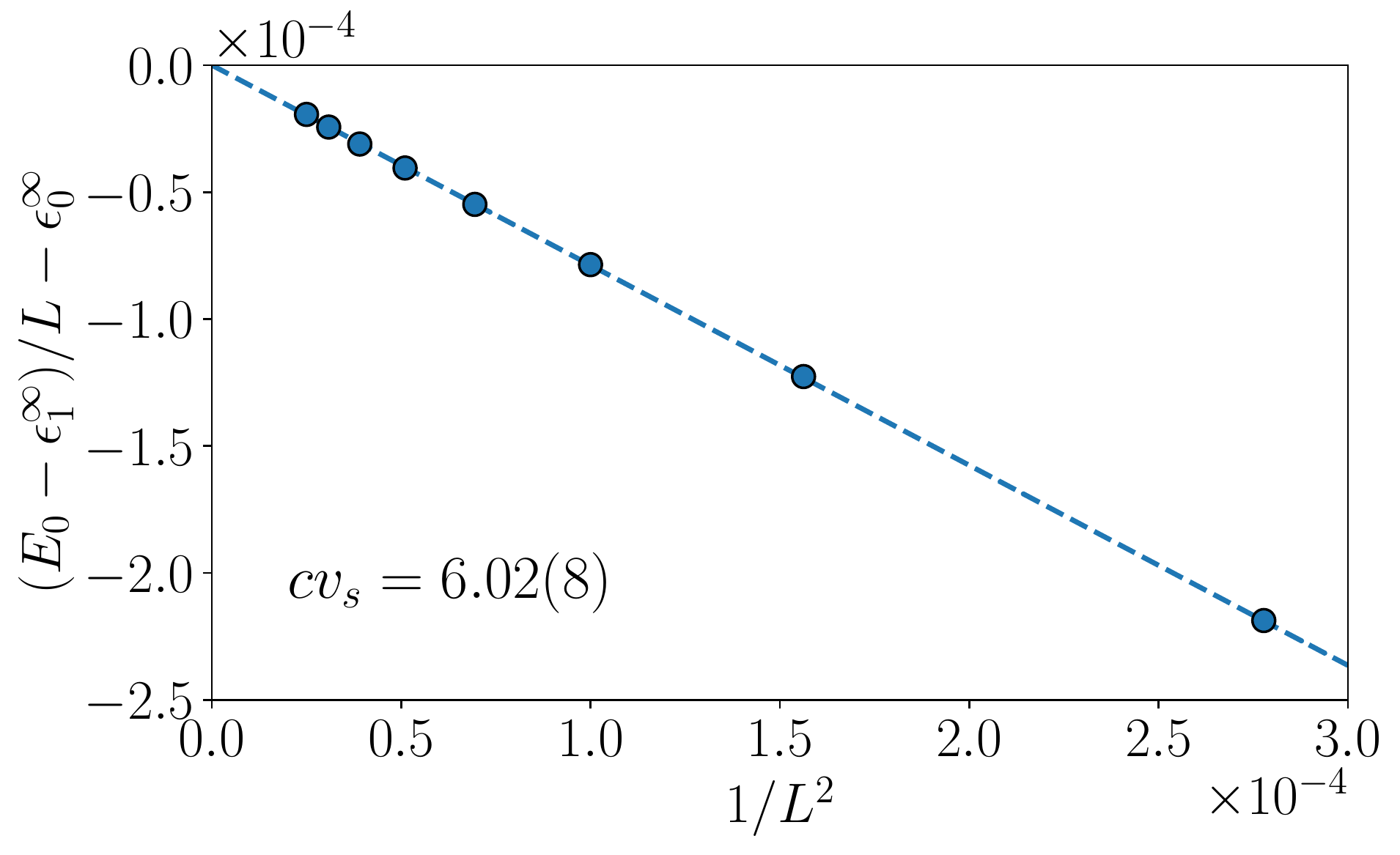}
  \caption{
    Finite-size scaling of the ground state energy $E_0$ at the critical point according to Eq.~\eqref{eq:gs_scaling} that gives $c v_s  \simeq 6$, where $c$ is the central charge and $v_s$ is the sound velocity.}
    \label{fig:gs_scaling}
\end{figure}

To estimate $v_s$, we perform the finite-size scaling of the ground state energy according to Eq.~\eqref{eq:gs_scaling} at the critical point (for similar scaling analysis, see \cite{Hallberg_1996, xavier_2010, Dalmonte_2012, Chepiga_2017}). The finite-size scaling (Fig.~\ref{fig:gs_scaling}) reveals 
\begin{equation}
c v_s = 6.02 \pm 0.08,
\label{eq:sound_velocity}
\end{equation}
confirming the Lorentz invariance of the critical point $(\mu_c^2/q^2 = 0.447, \lambda_c/q^2 = 0.0565)$ and the applicability of CFT framework.

\section{Numerical results}

Here we describe our numerical results that contribute to clarifying the nature of the field theory emerging at the critical point of the lattice model. 
For this, we set $\lambda/q^2 = 0.0565$ and  $\mu^2/q^2 = 0.447$ in the rest of the paper.

To reiterate, we employ DMRG simulations to extract the low-lying spectrum of the Hamiltonian~\eqref{eq:Hamil_pureM} with $U(1)$ symmetric MPS~\cite{singh_pra_2010,singh_prb_2011}. The MPS bond dimension used in our study is up to $\chi=600$, sufficient for convergence up to system sizes $120$. Each of the  bosonic species, `$a$' and  `$b$', has been truncated to $11$ levels, leading to the physical dimension of $121$ for the MPS tensors. Furthermore, in Sec.~\ref{sec:gaps}, we also simulate the full matter-gauge Hamiltonian~\eqref{eq:Hamil}. In that case, the MPS ansatz contains alternating tensors for matter and gauge degrees of freedom (see Fig.~\ref{fig:mps}). The gauge fields are then truncated to up to  15 levels in the electric basis. Details about the MPS based numerical calculations are presented in Appendix~\ref{app:mps}.

For the out-of-equilibrium analysis presented in Sec.~\ref{sec:spec_nonequil}, we use TDVP algorithms as described in~\cite{chanda20}. To reduce the computational complexity of TDVP runs, we truncate the bosonic species upto $8$ levels and the maximum MPS bond dimension is restricted upto $\chi=400$.

\subsection{The scaling of Renyi entanglement entropy}

The Renyi entanglement entropy of order $n$ for a block of $l$ sites is defined as
\begin{equation}
    S_{n}(l) = \frac{1}{1-n} \log \left( \text{Tr} \rho_l^n\right),
\end{equation}
where $\rho_l = \text{Tr}_{l+1, l+2, ..., L} \ket{\psi}\bra{\psi}$ is the reduced density matrix after tracing out the rest of the system. As limiting cases of the Renyi entropy, we get the von Neumann entropy as $S = \lim_{n \rightarrow 1} S_n$ and the entanglement ground state energy $\varepsilon_0 = \lim_{n \rightarrow \infty} S_n$, where $\varepsilon_0$ is the lowest eigenvalue of the entanglement Hamiltonian $H_l = -\log(\rho_l)$, i.e., $\varepsilon_0 = - \log \lambda_0$ where $\lambda_0$ is the smallest eigenvalue of the density matrix $\rho_l$.
It is to be noted that since we are integrating out the gauge fields, the resulting Hilbert space of matter has a well-defined tensor-product structure
and there are no ambiguities in defining the bipartitions associated with the gauge invariance (c.f.,~\cite{Casini_prd_2014, Ghosh2015, Soni2016}).

In a CFT, the finite-size scaling of the Renyi entanglement entropy of the bipartition of size $l$ in a chain with open boundary conditions and length $L$ is \cite{callan_geometric_1994, vidal_2003, calabrese_entanglement_2004}
\begin{equation}
S_n(l,L) = \frac{c}{12}(1 + \frac{1}{n}) W + b_n',
\label{eq:ent_scaling}
\end{equation}
where $c$ is the central charge of the corresponding CFT, $b_n'$ is a non-universal constant and $W$, the chord length, is a function of both $L$ and $l$:
\begin{equation}
W(l,L) =\log \left[   \frac{2 L}{\pi} \sin(\pi l /L)\right].
\end{equation}

\begin{figure}
    \centering
    \includegraphics[width=\linewidth]{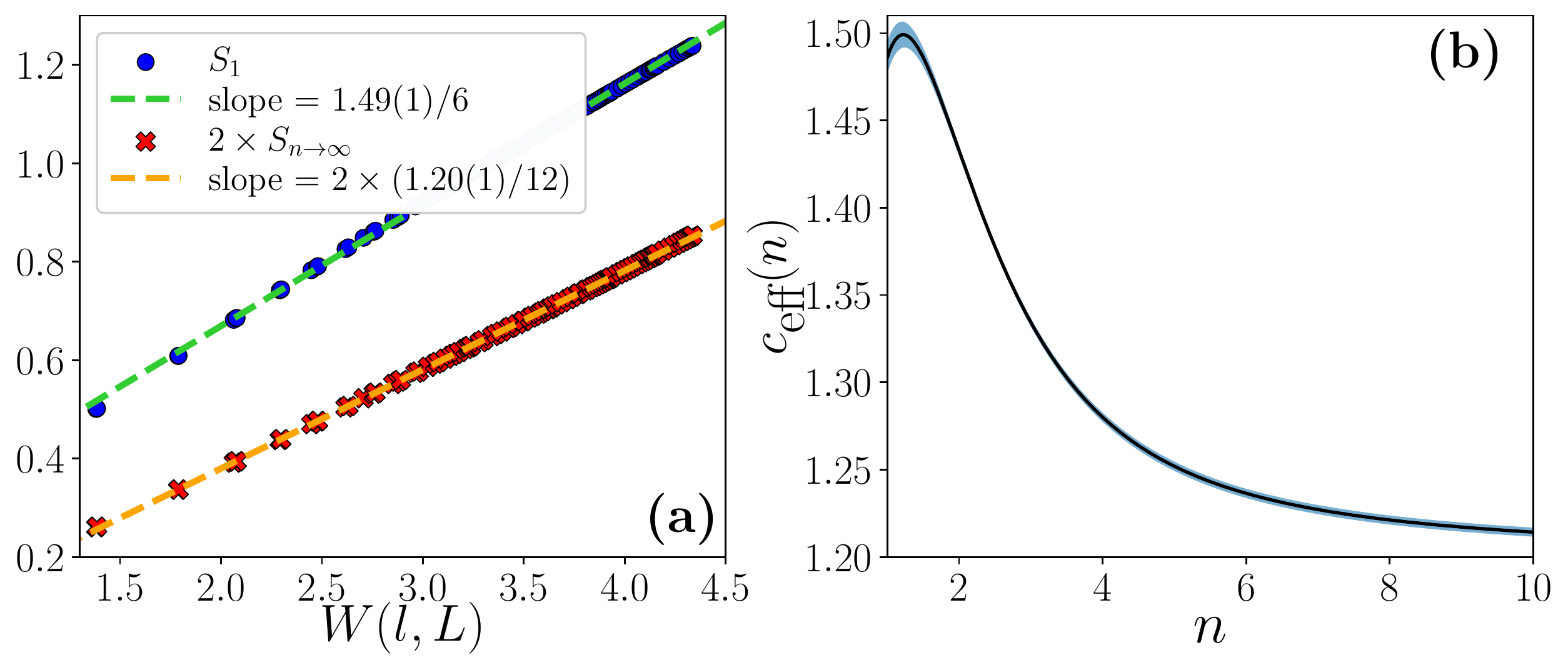}
    \caption{(Color online.)
    \textbf{(a)} The variation of the von Neumann entanglement entropy $S = S_{n \rightarrow 1}$ and the entanglement ground state energy $\varepsilon = S_{n \rightarrow \infty}$ as a function of the chord length $W$ according to Eq.~\eqref{eq:ent_scaling} at the critical SOQPT point.
    \textbf{(b)} The extracted values of the central charge after fitting the Renyi entropy data to Eq.~\eqref{eq:ent_scaling}. The bluish shade describes the error-bars in fitted values of $c_{\text{eff}}(n)$. For these plots, we have analyzed the data for system-sizes $L \in [40, 60, 80, 100, 120]$ at the critical point $(\lambda_c/q^2 = 0.0565, \mu_c^2/q^2 = 0.447)$.
    }
    \label{fig:central_charge_renyi}
\end{figure}

In Ref.~\cite{chanda2021}, using the above scaling of von Neumann entropy at the critical point, a central charge of $c=3/2$ ($c = 1.49(1)$ to be precise) has been extracted (see Fig.~\ref{fig:central_charge_renyi}\textbf{(a)}). It has been argued in \cite{chanda2021} that this value of the central charge can be explained as the sum of two different minimal models, each contributing to a piece of the total central charge,  a $c_f=1/2$ for a free Majorana fermion (the Ising sector) and a $c_b=1$ for a free boson. 

However, performing the same scaling for the entanglement ground state energy $\varepsilon_0$, the effective central
charge comes out to be $c_{\text{eff}}(n \rightarrow \infty) = 1.20(1)$ (see Fig.~\ref{fig:central_charge_renyi}\textbf{(a)}), which contradicts the value $c=3/2$. Moreover, on a closer inspection, we find that the values of central charge obtained by fitting the entropy data to Eq.~\eqref{eq:ent_scaling} is heavily dependent on the order $n$ as seen in Fig.~\ref{fig:central_charge_renyi}\textbf{(b)}).

\begin{figure}
    \centering
    \includegraphics[width=\linewidth]{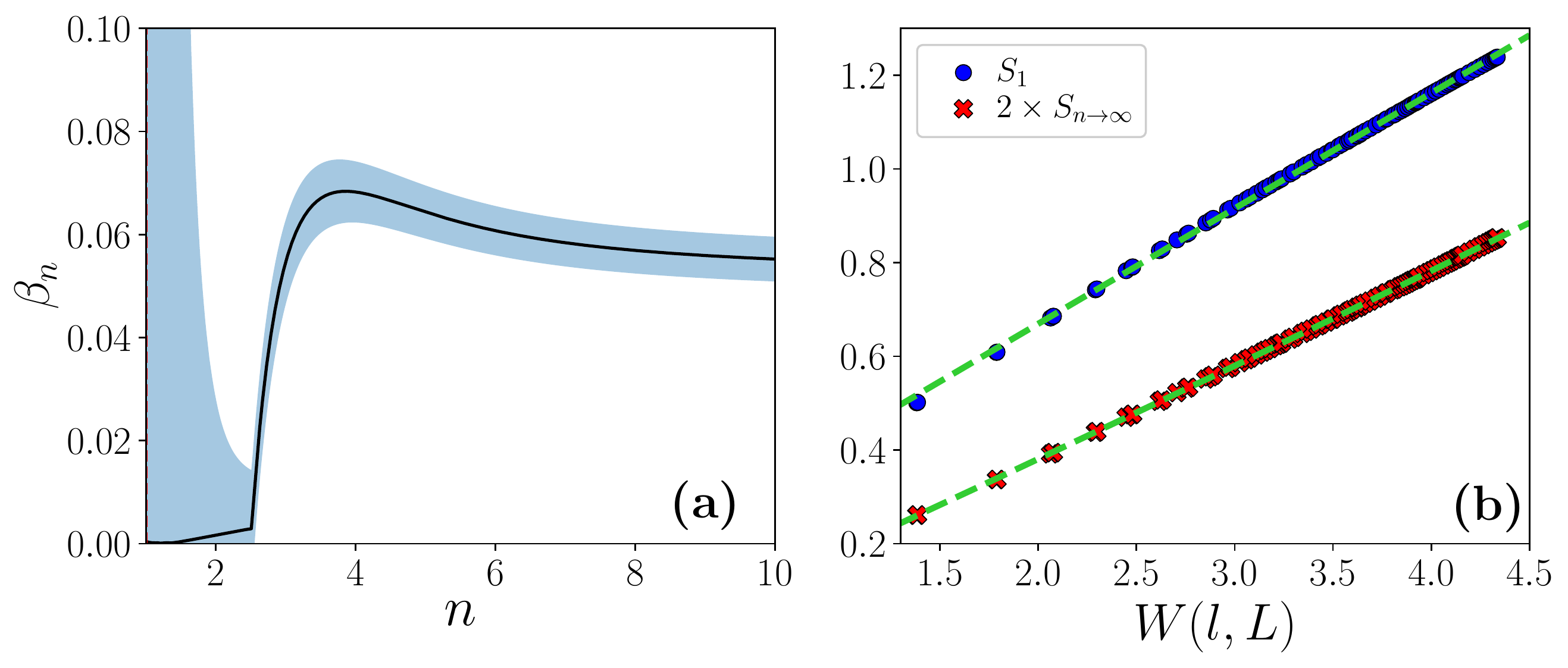}
    \caption{(Color online.)
    \textbf{(a)} The exponent $\beta_n$ in the finite-size correction to the entanglement entropy scaling (see Eq.~\eqref{eq:ent_scaling_2}). as a function of the Renyi order $n$. Here we fix $c=3/2$ (see text). The bluish shade describes the error bars in fitted values of $\beta_n$. Note that the error in the fitting is large when $c_{\text{eff}} \sim 3/2$ as the curve-fitting becomes ill-formed with redundant parameters within that regime.
    \textbf{(b)} The fitting of von Neumann entanglement entropy $S = S_1$ and the entanglement ground state energy $\varepsilon = S_{n \rightarrow \infty}$ including the correction term as in Eq.~\eqref{eq:ent_scaling_2}.
    Note that the entropic data (symbols) are exactly same as in Fig.~\ref{fig:central_charge_renyi}\textbf{(b)} -- the difference lies in the use of 
    Eq.~\eqref{eq:ent_scaling_2} in fitting these data.
    All other descriptions are the same as  those in Fig.~\ref{fig:central_charge_renyi}.
    }
    \label{fig:d_renyi_fit}
\end{figure}

Nevertheless, the analysis in Sec.~\ref{sec:sound_velocity} and in the following Sec.~\ref{sec:spec_nonequil} suggest that the central charge of the system is indeed compatible with $c=3/2$. 
In Sec.~\ref{sec:sound_velocity}, by utilizing the finite-size scaling of ground state energy we have shown that $c v_s \simeq 6$, whereas using the out-of-equilibrium local quench spectrocopy in the following Sec.~\ref{sec:spec_nonequil} we verify that the sound velocity is consistent with $v_s=4$ -- thereby leaving $c=3/2$ as the only possibility. These results are obviously independent of the entropy scaling.

Here, we try to explain the discrepancies in the scaling of different Renyi entropies by finite-size corrections in the scaling law of Eq.~\eqref{eq:ent_scaling}. Specifically, it has been shown in~\cite{cardy_unusual_2010} that due to irrelevant and marginally irrelevant bulk operators in the Hamiltonian, the Renyi entropies can attain corrections in finite systems, where the correction terms can depend on $n$ and on the scaling dimensions of such bulk operators (see also \cite{Eriksson2011, Laflorencie2006}). Since the numerical determination of the scaling dimensions of such irrelevant (or marginally irrelevant) bulk operators can be very non-trivial, here we consider a generic correction term, having the same form as in \cite{cardy_unusual_2010}, to the scaling formula of Eq.~\eqref{eq:ent_scaling} as follows:
 \begin{equation}
    S_n(l,L) = \frac{c}{12}(1 + \frac{1}{n}) W + b_n' + d_n \left( \exp(W) \right)^{-\beta_n},
    \label{eq:ent_scaling_2}
\end{equation}
where $d_n$ and $\beta_n$ are $n$-dependent quantities to be determined.
It is to be noted that since Eq.~\eqref{eq:ent_scaling_2} has four independent parameters, the curve fitting process is underconstraint for the given entropic data leading to unfaithful extraction of these parameters. Because of this, we fix the central charge to $c=3/2$, verified independently from the ground-state energy scaling in Sec.~\ref{sec:sound_velocity} (see Eq.~\eqref{eq:sound_velocity}) and the local quench spectroscopy in Sec.~\ref{sec:spec_nonequil}, when fitting fitting Eq.~\eqref{eq:ent_scaling_2} to the entropic data.

The variation of the exponent $\beta_n$ as a function of the order $n$ is shown in Fig.~\ref{fig:d_renyi_fit}\textbf{(a)} along with its errors for fixed $c = 3/2$. For small $n$, corrections to the original formula are small (the corresponding error is not informative as such term does not contribute to the fit). Oppositely, for large $n$, where we observe $c_{\text{eff}} \not \approx 3/2$ (Fig.~\ref{fig:central_charge_renyi}\textbf{(b)}), corrections are indeed strong and the coefficient $\beta_n$ is essentially $n$-independent. This strongly confirms the fact that extracting the central charge from such large-$n$ Renyi entropies is not reliable using the original scaling relation when finite-size effects are strong.  Fig.~\ref{fig:d_renyi_fit}\textbf{(b)} shows the actual entropy scalings for the von Neumann entanglement entropy $S = S_1$ and the entanglement ground state energy $\varepsilon = S_{n \rightarrow \infty}$ according to Eq.~\eqref{eq:ent_scaling_2} with fixed $c=3/2$.

\subsection{Spectral analysis at equilibrium}

To detect and characterize the Majorana fermionic and the bosonic part of the theory, we first analyze the energies  of the ground state and low-lying excited states and their finite-size scaling at the critical point $(\lambda_c/q^2 = 0.0565, \mu_c^2/q^2 = 0.447)$ obtained from DMRG simulations.

\subsubsection{The neutral gap and the charge gap}
\label{sec:gaps}

We calculate the neutral gap ($\Delta E_{N}$) and the charge gap ($\Delta E_C$) at the critical point. The gaps are defined as
\begin{align}
&\Delta E_{N} = E_1(Q=0) - E_0(Q=0), \nonumber \\
&\Delta E_{C} = E_0(Q = 1) + E_0(Q = - 1) - 2 E_0(Q=0).
\end{align}
where $E_n(Q)$ represents the $n^{th}$ energy eigenvalue at the $Q = \braket{\hat{Q}}$ quantum sector extracted by means of excited state DMRG.

\begin{figure}
    \centering
    \includegraphics[width=\linewidth]{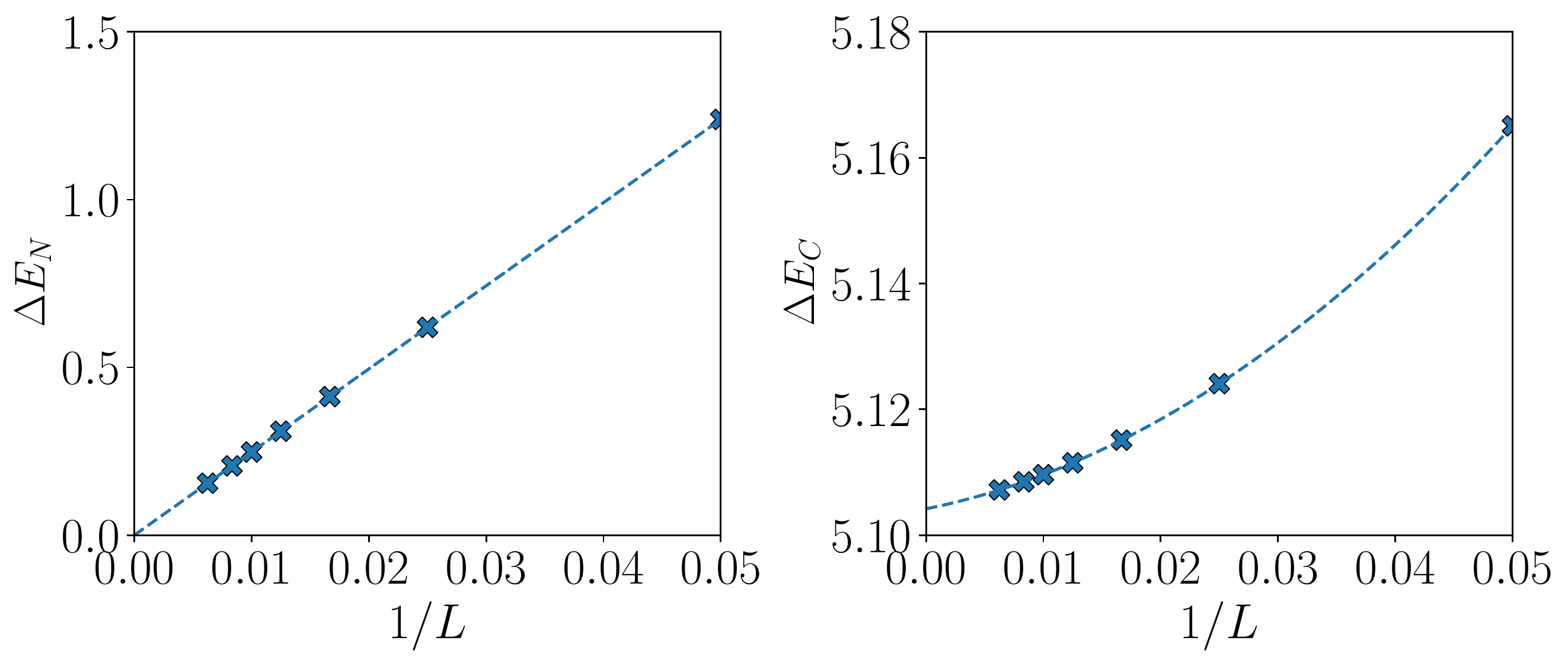}
    \caption{(Color online.)
    The scalings of \textbf{(a)} the neutral gap $\Delta E_N$ and \textbf{(b)} the charge gap $\Delta E_C$ with the system-size $L$ at the critical SOQPT point. The neutral gap decays to zero in the thermodynamic limit as $\sim 1/L$, while the charge gap remains finite and its thermodynamic value ($5.104(4)$) is extracted by fitting a quadratic function in $1/L$.
    }
    \label{fig:gaps}
\end{figure}

As expected, the neutral gap $\Delta E_{N}$ decays to zero in the thermodynamic limit as $\sim 1/L$ (see Fig.~\ref{fig:gaps}\textbf{(a)}). On the other hand, quite interestingly, the charge gap $\Delta E_C$ saturates to a finite value $\Delta E_{C} = 5.104(4)$ in the thermodynamic limit (see Fig.~\ref{fig:gaps}\textbf{(b)}). This happens  due to the gauge invariance of the system, as a single charge excitation must be accompanied by a string of electric fields (semi-infinite in the thermodynamic limit) whose energy cannot vanish.

\begin{figure}
    \centering
    \includegraphics[width=0.8\linewidth]{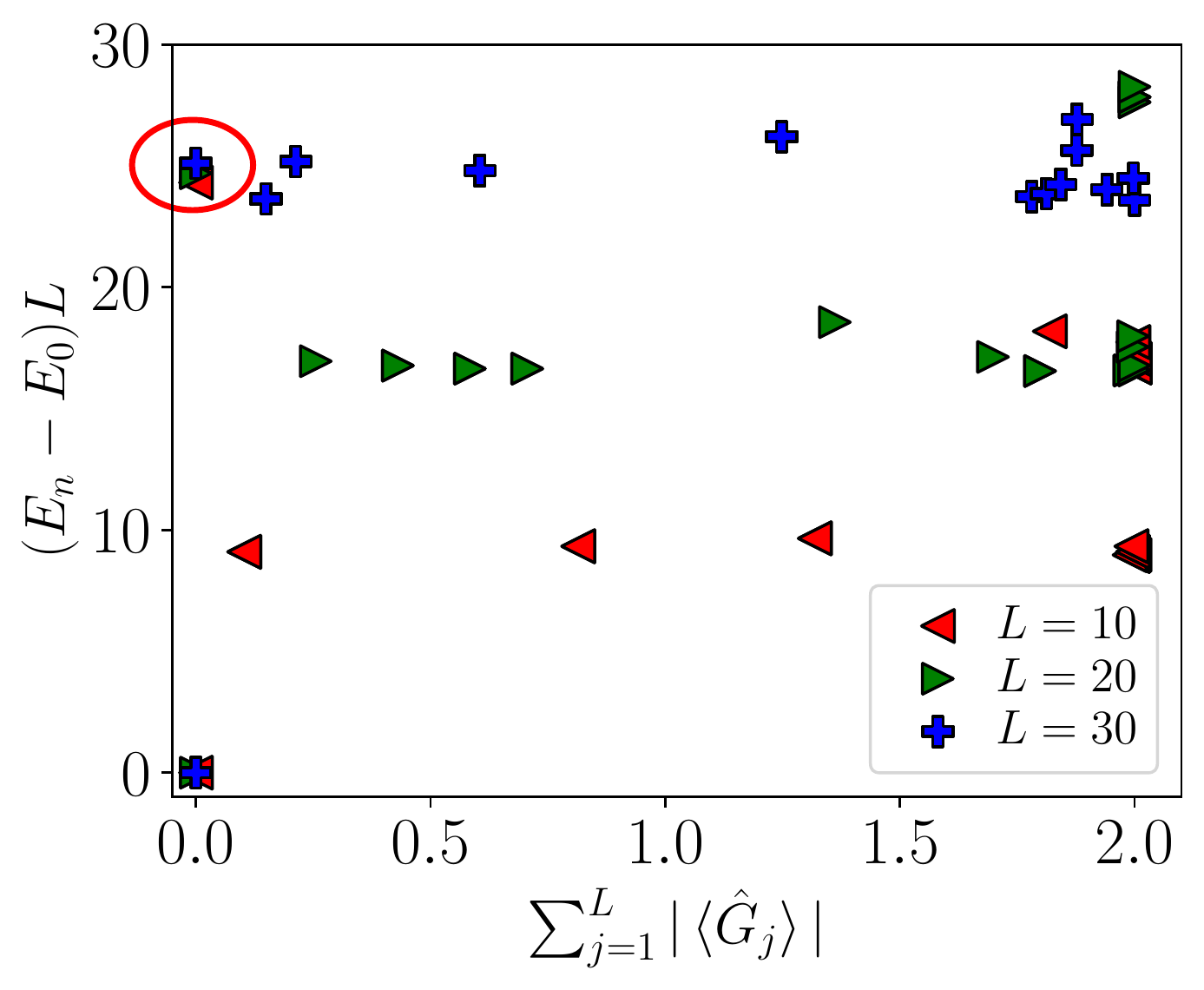}
    \caption{(Color online.)
    Energy gaps of few excited states obtained from excited-state DMRG when we do not enforce Gauss law by integrating out the gauge fields, and instead treat the Hamiltonian \eqref{eq:Hamil} as it is with conserved total charge $Q = 0$.
    Here, we truncate the gauge fields to 15 electric levels. The x-axis, $\sum_{j=1}^L|\braket{\hat G_j}|$, denotes the deviation from the Gauss law $\braket{\hat G_j} = 0$ for all $j = 1, \cdots, L$.
    The red circle demarcates the first neutral excitations that obey the Gauss law.
    Note here that the excited state DMRG is unable to find all the low-lying states coming from different gauge sectors, specifically when $\sum_{j=1}^L|\braket{\hat G_j}| \neq 0$. Nevertheless, we can see the existence of many low-lying states coming from different gauge sectors in the low-energy spectrum.
    }
    \label{fig:gapsNoGauss}
\end{figure}

Therefore, by imposing the Gauss law, we are throwing away a large chunk of the low-energy spectrum. This is illustrated in Fig.~\ref{fig:gapsNoGauss}, where we show that there are indeed many low-lying states that come from other gauge sectors (i.e., non-zero background fields) by performing the excited state DMRG calculations for the matter-gauge Hamiltonian \eqref{eq:Hamil}. It is to be noted that since gauge invariance cannot be broken spontaneously, excited state DMRG cannot scan all possible low-lying gauge sectors, and its reliability in iteratively finding excited states one-after-another degrades very quickly when we do not impose the Gauss law.

\subsubsection{Conformal towers}
\label{sec:conf_tower}

\begin{figure}
    \centering
    \includegraphics[width=0.85\linewidth]{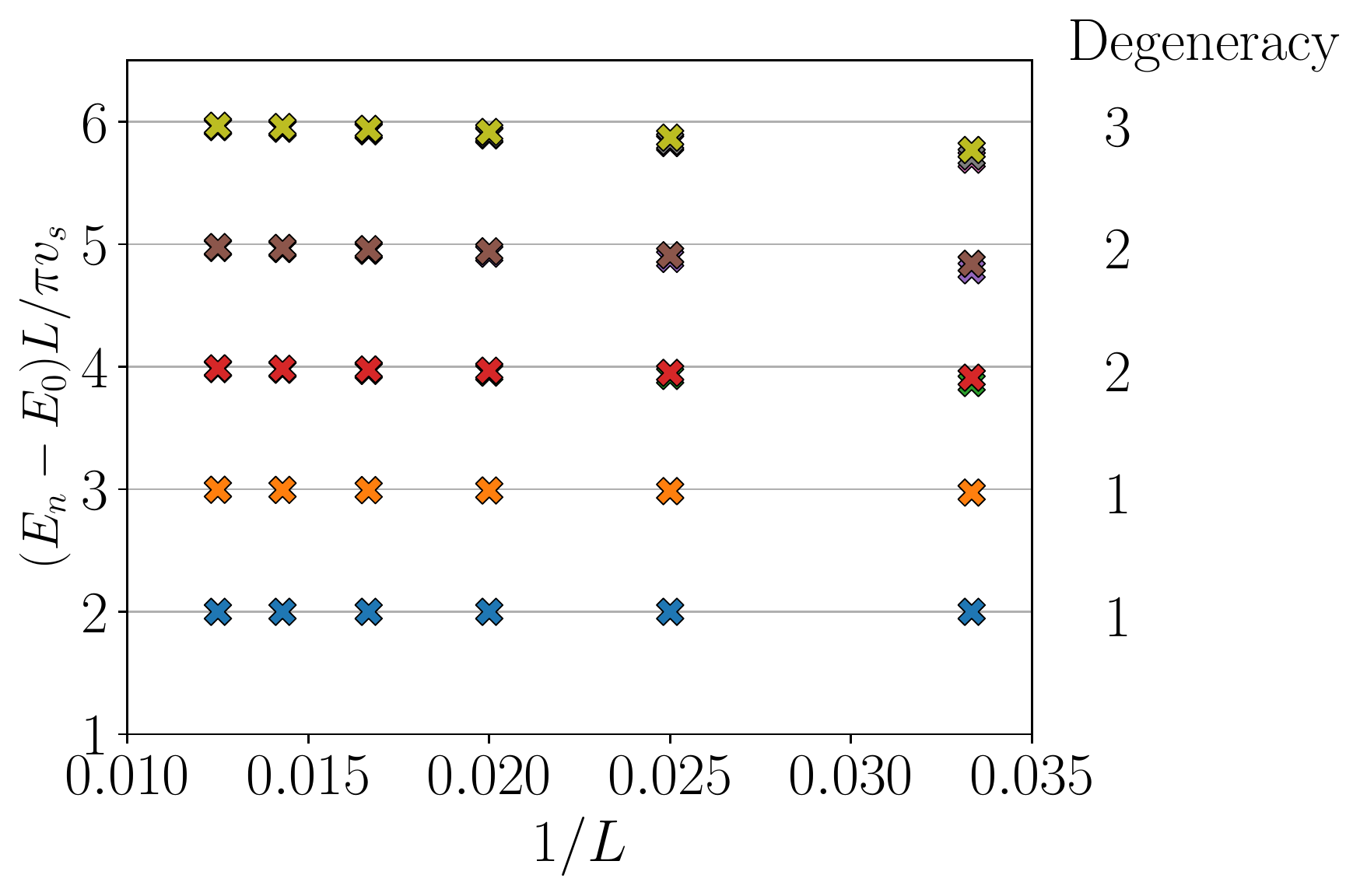}
    \caption{(Color online.)
    First few boundary scaling dimensions $x_n = (E_n - E_0)L/\pi v_s$ in the conformal tower of the critical point using excited-state DMRG. 
    }
    \label{fig:gaps_q0}
\end{figure}

Here, we analyze the excitation spectra of the system at a critical point using the predictions of CFT by using the excited-state DMRG. The excited state energies vary with the system-size as
\begin{equation}
    E_n(L) = E_0(L) + x_n \frac{\pi v_s}{L},
\end{equation}
where $x_n$ are boundary scaling dimensions, organized in conformal towers \cite{cardy1986,cardy1986a}. In Fig.~\ref{fig:gaps_q0}, we plot $x_n = (E_n - E_0)L/\pi v_s$ for first $9$ excited states along with their degeneracies obtained from the excited-state DMRG calculations. The boundary exponents that we get from Fig.~\ref{fig:gaps_q0} are  $x_n = [2, 3, 4, 5, 6, ...]$ with degeneracies $[1, 1, 2, 2, 3,...]$. 
These values of the boundary exponents and corresponding degeneracies match with the CFT prediction for the critical Ising model with fixed boundary condition~\cite{rochacaridi_1985, Evenbly2014, Chepiga_2017}. This structure of the conformal tower confirms the Ising or free Majorana fermion part of the critical point. On the other hand, we completely miss the bosonic part of the spectrum at the critical point. However, we can make 
an educated guess based on the results of Sec.~\ref{sec:gaps} (specifically, of Fig.~\ref{fig:gapsNoGauss}) that the bosonic excitations, that carry charges, might be hidden in other gauge sectors.

\begin{figure}
    \centering
    \includegraphics[width=\linewidth]{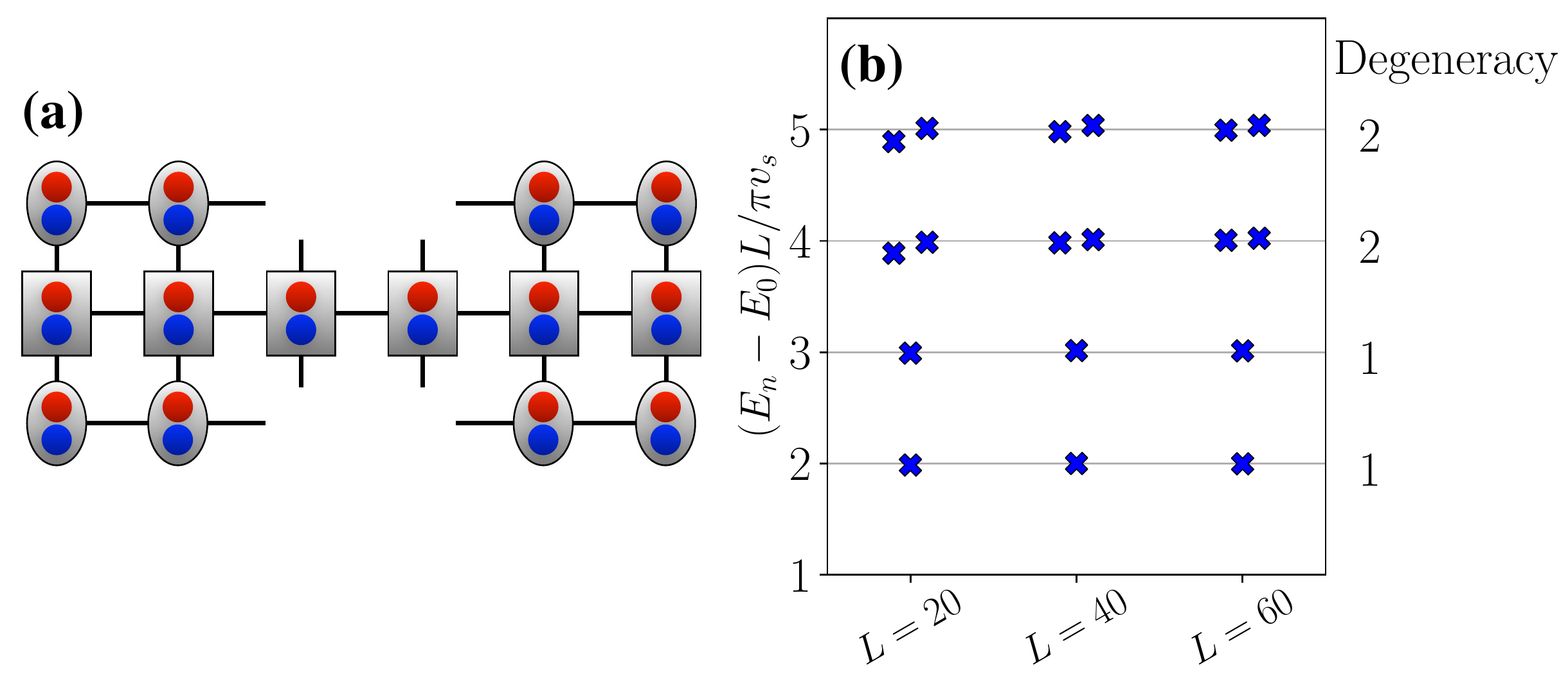}
    \caption{(Color online.) \textbf{(a)} The graphical representation of the mid-chain effective Hamiltonian constructed from the MPO and MPS tensors to target the ground state of the  Hamiltonian~\eqref{eq:Hamil_pureM}. 
    \textbf{(b)} The excitation spectra of the effective Hamiltonian depicted in \textbf{(a)} at the critical point.
    }
    \label{fig:effec_spec}
\end{figure}

To validate this result further, we analyze the spectra of the effective Hamiltonian constructed to target the ground states of the Hamiltonian \eqref{eq:Hamil_pureM} during DMRG sweeps (see Fig.~\ref{fig:effec_spec}\textbf{(a)}). In Ref.~\cite{Chepiga_2017}, it has been shown that at the critical point, the low-lying eigenstates of the effective Hamiltonian, within the framework of the standard ground state algorithm, gives correct excitation spectra of the original Hamiltonian. In Fig.~\ref{fig:effec_spec}\textbf{(b)} we show the first few eigenstates of the effective Hamiltonian at the critical point. Once again, 
the extracted boundary exponents and corresponding degeneracies match with the Isinf CFT.

\subsection{Determination of the Luttinger parameter}

\begin{figure}
    \centering
    \includegraphics[width=\linewidth]{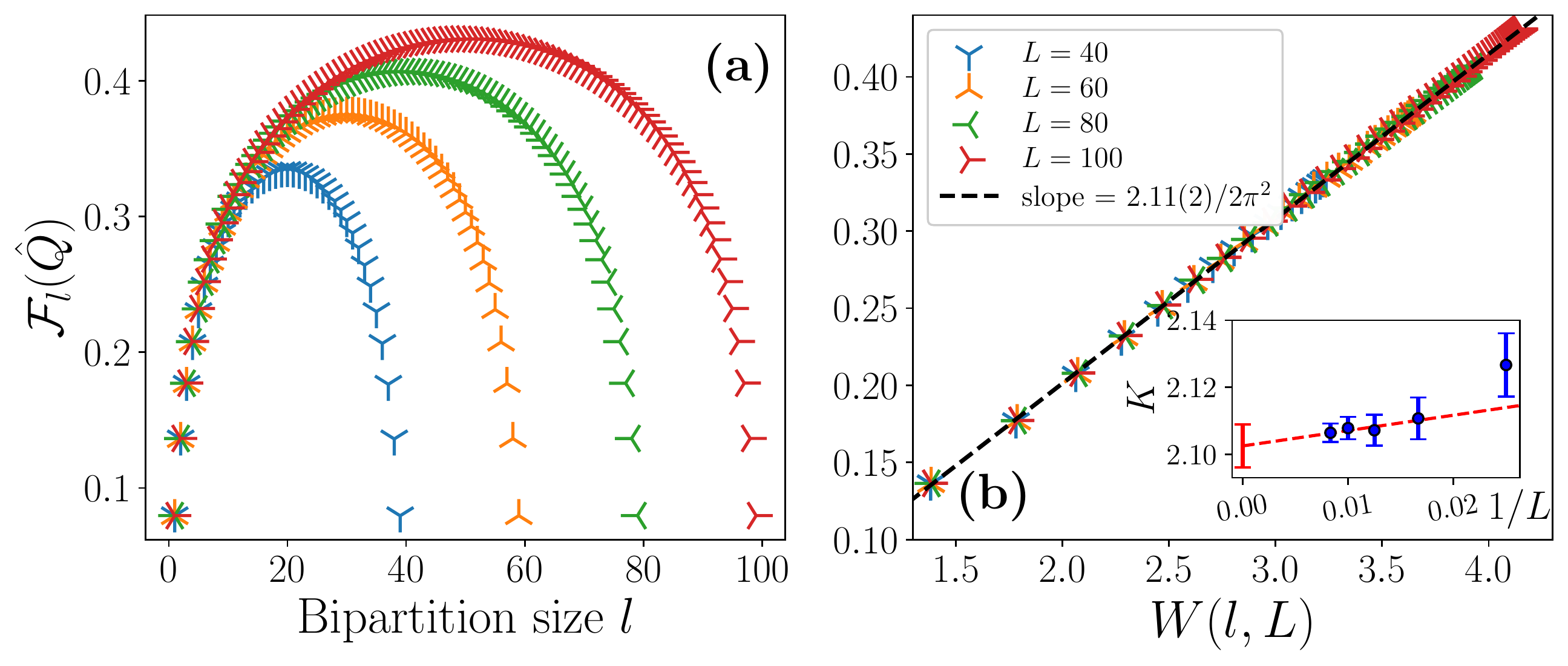}
    \caption{(Color online.) The scaling of local fluctuations $\mathcal{F}_l(\hat{Q})$ in relation to the chord length $W(l, L)$.
\textbf{(a)} The pattern of local fluctuations with respect to the bipartition size for different system sizes is plotted and indicated in the figure.
\textbf{(b)} The fit performed with the data for $L \in [40, 60, 80, 100, 120]$, based on Eq.~\eqref{eq:luttingerKscaling}, suggests that Luttinger parameter is  $K = 2.11(2)$. The inset shows the negligible dependence of $K$ on the system size. To estimate Luttinger parameter $K$ at the thermodynamic limit, we extrapolate the extracted values of $K$ for different system-sizes $L \in [60, 120]$ using a linear fit in $1/L$, yielding $K = 2.10(1)$.
}
    \label{fig:LuttinterK}
\end{figure}

Since we could not characterize the bosonic part of the critical theory from the low-energy excitation spectrum, 
we take the route of determining the Luttinger parameter $K$ associated with the free bosonic part that is described by the Tomonaga-Luttinger liquid theory \cite{Giamarchi_book, Gogolin2004}. Here, we extract $K$ from the scaling of the bipartite fluctuations as described in Refs.~\cite{Song10, Song12, Rachel12}, under the assumption that charge excitations are indeed those responsible for the emergence of the $c_{b}=1$ sector of theory. It has been shown that in a Tomonaga-Luttinger liquid,
for a global $U(1)$ conserved quantity $\mathcal{O}$, the local fluctuations
\begin{equation}
    \mathcal{F}_l (\mathcal{O}) = \Braket{\left(\sum_{j\leq l} \mathcal{O}_j \right)^2} - \Braket{\sum_{j \leq l} \mathcal{O}_j}^2
\end{equation}
obey the scaling of the form
\begin{equation}
    \mathcal{F}_l (\mathcal{O}) = \frac{K}{2 \pi^2} W(l, L) + \text{const.}
    \label{eq:luttingerKscaling}
\end{equation}
In the present scenario, the global conserved quantity is the total dynamical charge $\hat Q$. Using the local gauge invariance, we can also rewrite the local fluctuations of the dynamical charge as
\begin{equation}
    \mathcal{F}_l(\hat Q) = \Braket{\hat{L}_l^2} - \Braket{\hat{L}_l}^2,
\end{equation}
where we have used the fact that $\hat{L}_{l} = \sum_{j\leq l} \hat Q_j$. In Fig.~\ref{fig:LuttinterK}, we present the variations of local fluctuations as a function of both bipartition and system size. This figure provides evidence that the Luttinger parameter is $K = 2.10(1)$, which characterizes the existence of a strongly interacting bosonic component in the theory. 

\subsection{Spectral analysis from out-of-equilibrium dynamics}
\label{sec:spec_nonequil}

We now analyze the critical point by considering the response of the system to a local quench as suggested in ~\cite{villa2019, villa2020a}. For that, we create a local excitation at the middle of the chain using the operator
\begin{equation}
\hat{\mathcal{M}}^Q = \hat{\phi}^{\dagger}_{L/2}\hat{\phi}_{L/2+1}
\label{eq:charge_op}
\end{equation}
that will create $Q=\pm 1$ dynamical charges at sites $L/2$ and $L/2 +1$ respectively on top of the ground state $\ket{\Omega}$, such that our initial state becomes
\begin{equation}
    \ket{\psi^Q}(t=0) = \mathcal{N} \hat{\mathcal{M}} \ket{\Omega},
    \label{eq:init_state_Q}
\end{equation}
where $\mathcal{N}$ is a normalization constant. Since, the operator $\hat{\mathcal{M}}^Q$ is a local element of the Hamiltonian \eqref{eq:Hamil_pureM}, we can expect that the overlaps of $\ket{\psi^{Q}}$ with low-lying excited states are non-vanishing.

For any local observable $\hat{\mathcal{O}}_j$, we analyze the spectral properties by means of the (discrete) Fourier transform:
\small
\begin{equation}
    \mathcal{F}_{\mathcal{O}}(k, \omega) = \frac{2 \pi}{L T} \delta t \sum_{j=1}^{L} e^{-i k (j-\frac{L}{2})} \sum_{n=0}^{t_N} e^{-i \omega t_n} \left(\braket{\mathcal{O}_j}(t_n) - \braket{\mathcal{O}_j}_{\Omega}\right),
    \label{eq:fourier}
\end{equation}
\normalsize
where $t_n = n \delta t$ is the discrete time-steps, $T = t_N \delta t$ is the total time of the evolution, and $\braket{\mathcal{O}_j}_{\Omega}$ defines the expectation value of $\hat{\mathcal{O}}_j$ with respect to the the ground state $\ket{\Omega}$ (see~\cite{paeckel_aop_2019} for details).

Since the neutral gap vanishes while the charge gap remains finite at the critical point, we expect that  for neutral operators such as $\hat{\phi}^{\dagger}_j \hat{\phi}_j$, $\hat{\Pi}^{\dagger}_j \hat{\Pi}_j$, or $\hat{L}^2_j = (\sum_{l \leq j} \hat{Q}_l)^2$ the Fourier transform will provide a gapless dispersion with $\omega(k) \approx v_s |k|$ for $k \ll 1$, and 
for any charge carrying operators like $\hat{Q}_j$ or $\hat{L}_j = \sum_{l \leq j} \hat{Q}_l$, the $\mathcal{F}_{\mathcal{O}}(k, \omega)$ will give a gapped spectrum~\cite{villa2019, villa2020a}.

\begin{figure}[thb]
\centering
\includegraphics[width=\linewidth]{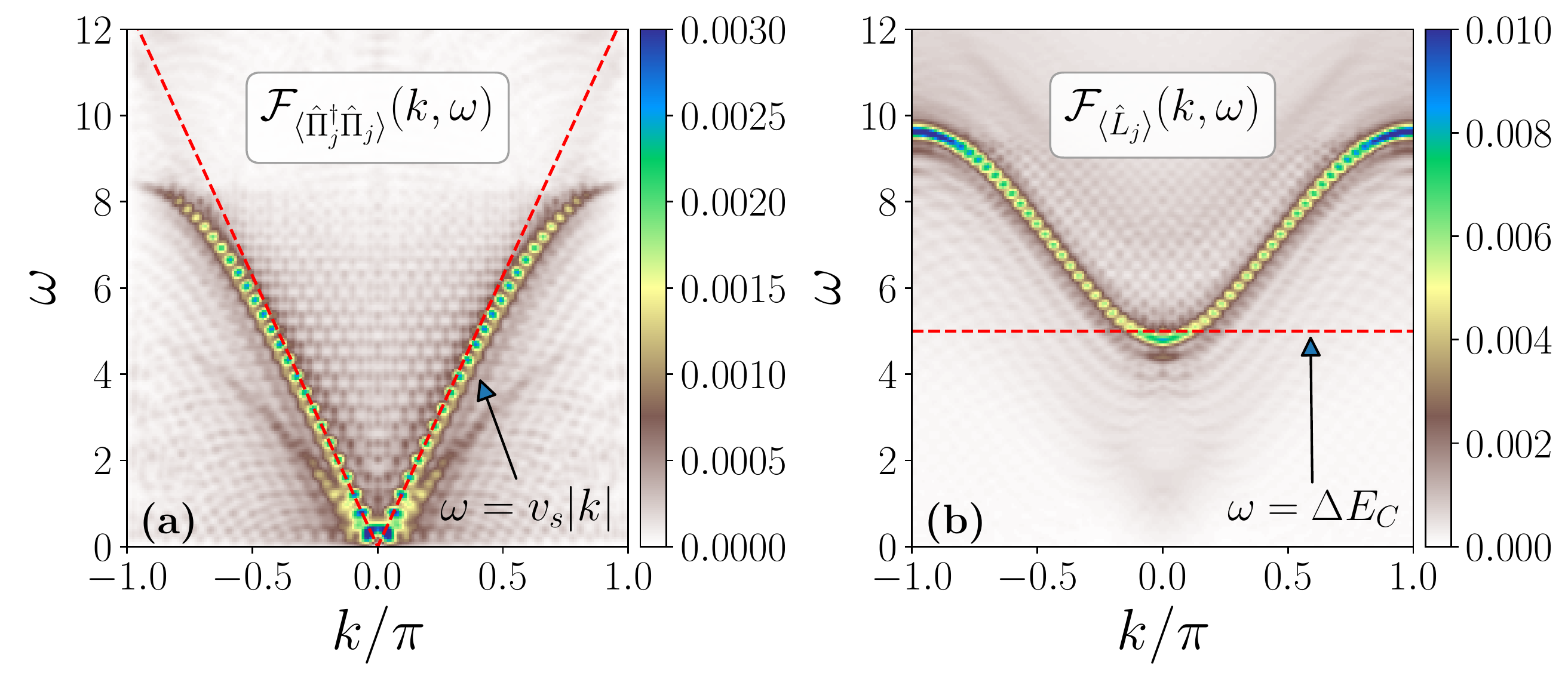}
\caption{(Color online.) The dispersion relations from the Fourier transform (Eq.~\eqref{eq:fourier}) of the local operators \textbf{(a)} $\hat{\Pi}_j^{\dagger} \hat{\Pi}_j$ and \textbf{(b)} $\hat{L}_j$ at the critical point for the initial state \eqref{eq:init_state_Q} for a system of size $L=60$. Here we apply the operator $\hat{\mathcal{M}}^Q$ (see Eq.~\eqref{eq:charge_op}) to excite the ground state.
\label{fig:dispersion_crit_Q}
}
\end{figure}

From the Fourier transform of the neutral operators, e.g., $\mathcal{F}_{\braket{\hat{\Pi}^{\dagger}_j \hat{\Pi}_j}}(k, \omega)$ in Fig.~\ref{fig:dispersion_crit_Q}\textbf{(a)}, we are getting the gapless spectrum as expected. The signal in the Fourier transform matches with 
$v_s \approx 4$ (red dashed line in Fig.~\ref{fig:dispersion_crit_Q}\textbf{(a)}) extracted from the scaling of ground state energy (see Sec.~\ref{sec:sound_velocity}).
On a careful inspection, we find another faint signal corresponding to $v_s \approx 2.4$. However, this value of the sound velocity is not consistent with earlier equilibrium analysis coming from the scaling of ground state energy (Eq.~\eqref{eq:sound_velocity}). 
This faint signal might be a spurious effect coming from numerical errors, as also seen in \cite{paeckel_aop_2019}. There is also a possibility that this might come from the dispersion in $Q=\pm 1$ sectors as the excitation performed by the operator $\hat M^Q$ couples these sectors to the present $Q=0$ one. 
On the other hand, in Fig.~\ref{fig:dispersion_crit_Q}\textbf{(b)}, we see that the dispersion relation from $\mathcal{F}_{\braket{\hat{L}_j}}(k, \omega)$ is gapped and the gap is exactly the same as the charge gap $\Delta E_C$.

To be sure that there is only one sound velocity at $v_s=4$, we perform two other forms of time-evolution in such a way that the local excitation now does not create any dynamical charges so that we can solely examine the neutral excitations. Specifically, we excite the ground state $\ket{\Omega}$ with the neutral operators
\begin{align}
    &\hat{\mathcal{M}}^1 = \hat{\phi}^{\dagger}_{L/2}\hat{\phi}_{L/2}\hat{\phi}^{\dagger}_{L/2+1}\hat{\phi}_{L/2+1} \label{eq:chargeless_op1} \\
&\hat{\mathcal{M}}^2 = \hat{\phi}^{\dagger}_{L/2}\hat{\phi}_{L/2} + \hat{\phi}^{\dagger}_{L/2+1}\hat{\phi}_{L/2+1}, \label{eq:chargeless_op2} 
\end{align}
and then perform a similar Fourier analysis as before.

\begin{figure}
\centering
\includegraphics[width=\linewidth]{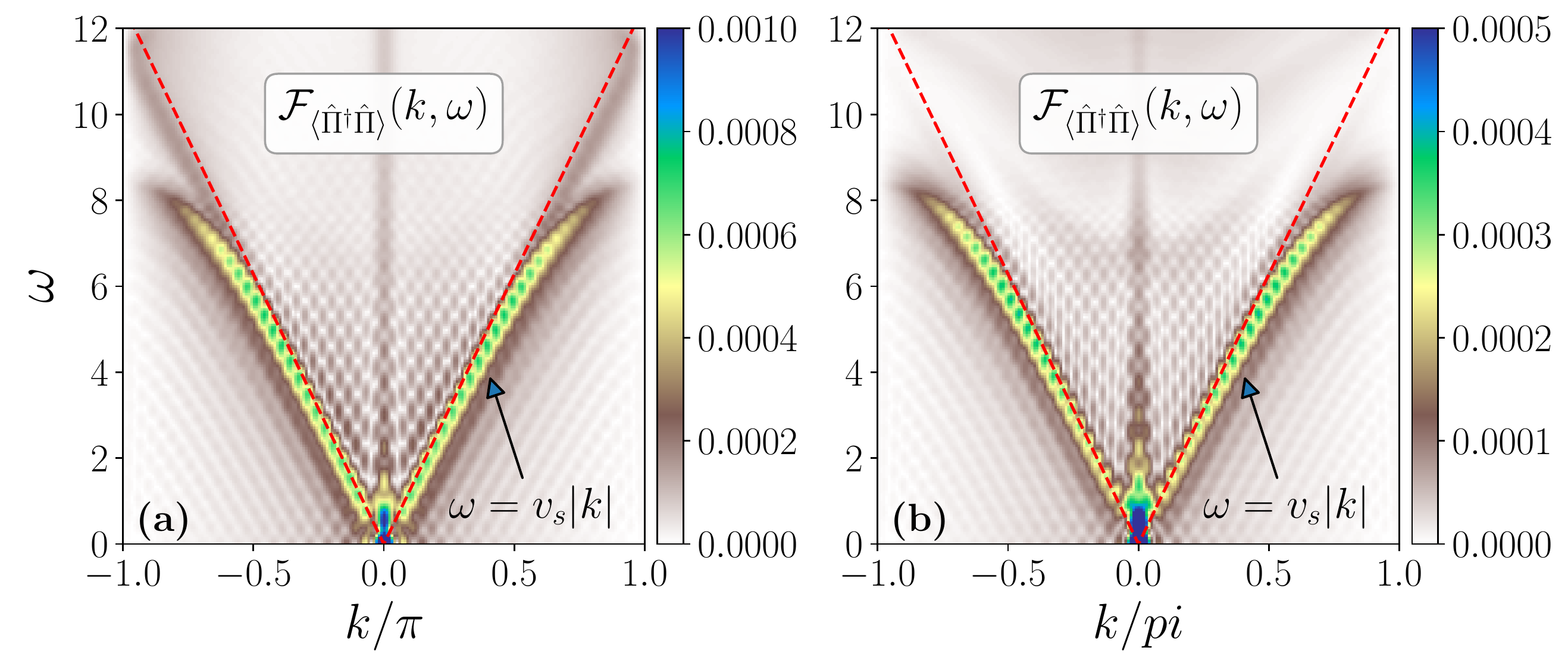}
\caption{(Color online.) Same as in Fig.~\ref{fig:dispersion_crit_Q}\textbf{(a)} but for the initial states excited with the operators given in \textbf{(a)} Eq.~\eqref{eq:chargeless_op1} and  \textbf{(b)} Eq.~\eqref{eq:chargeless_op2}.}
\label{fig:NoCharge_crit_disp}
\end{figure}

In Fig.~\ref{fig:NoCharge_crit_disp}, we show the Fourier transforms of $\hat{\Pi}_j^{\dagger} \hat{\Pi}_j$ for such initial states. Clearly, there is only one sound velocity of $v_s = 4$ that we can observe.

\section{Conclusions}

In this article, we presented an in-depth analysis of the critical properties of the 1+1D Abelian-Higgs model on a discretized lattice. Our approach provides a comprehensive understanding of the model's behavior, including its bosonic and fermionic components.

First, we addressed the ambiguity surrounding the value of the central charge at the critical point. We confirmed that the central charge is $c=3/2$, which can be interpreted as a direct sum of gapless compactified bosons with $c_b=1$ and critical Majorana fermions with $c_f=1/2$. We independently corroborated this value from both equilibrium and out-of-equilibrium analysis. Our results also reveal that the discrepancy in extracting the central charge from different Renyi entropies can be attributed to finite size effects coming from irrelevant or marginally irrelevant bulk operators.

To identify the bosonic and fermionic components of the model, we have used a two-fold approach. Firstly, we find that the excitation spectrum of the model is identical to the one of the critical Ising chain with fixed boundary conditions, i.e., the $c=1/2$ free fermionic part. Quite interestingly, we find no sign of the bosonic excitations in the gauge-invariant spectrum. Our examination of the excitation spectrum in different superselection sectors leads us to make an informed guess that the bosonic excitation spectrum may be `hidden' in other gauge sectors. We have then determined the Luttinger parameter, which confirms the existence of a gapless Tomonaga-Luttinger liquid with a value of Luttinger parameter $K \approx 2.1$.

The fact that the bosonic spectrum that contributes to the entanglement entropy at the critical point could be hidden in different superselection sectors suggests that a better understanding of the boundary CFT involved with the computation of the entanglement entropy here could be necessary. Also, some further hints could be obtained from the analysis of the symmetry-resolved entanglement spectrum that will be performed elsewhere.

Finally, we analyzed the local quenched dynamics of the critical system. By performing spectroscopy via discrete Fourier transform, we extract the dispersion relations for the interacting system. Our analysis reveals that there exists only one speed of sound in this gauge invariant system, which is consistent with our equilibrium analysis.

\appendix

\section{Details about the matrix product states (MPS) simulations}
\label{app:mps}

In this paper, we use matrix product states (MPS)\cite{schollwock_aop_2011, orus_aop_2014} based tensor network simulations to obtain the results. Specifically, to extract the ground state and low-lying excited states, we use a strictly single-site variant of density matrix renormalization group (DMRG) algorithm~\cite{white_prl_1992,white_prb_1993,white_prb_2005,schollwock_rmp_2005}  with subspace expansion \cite{Hubig_prb_2015}. For the out-of-equilibrium real-time dynamics, we employ the time-dependent variational principle (TDVP) method~\cite{haegeman_prl_2011, haegeman_prb_2016, koffel_prl_2012, paeckel_aop_2019}.

\begin{figure}
\centering
\includegraphics[width=\linewidth]{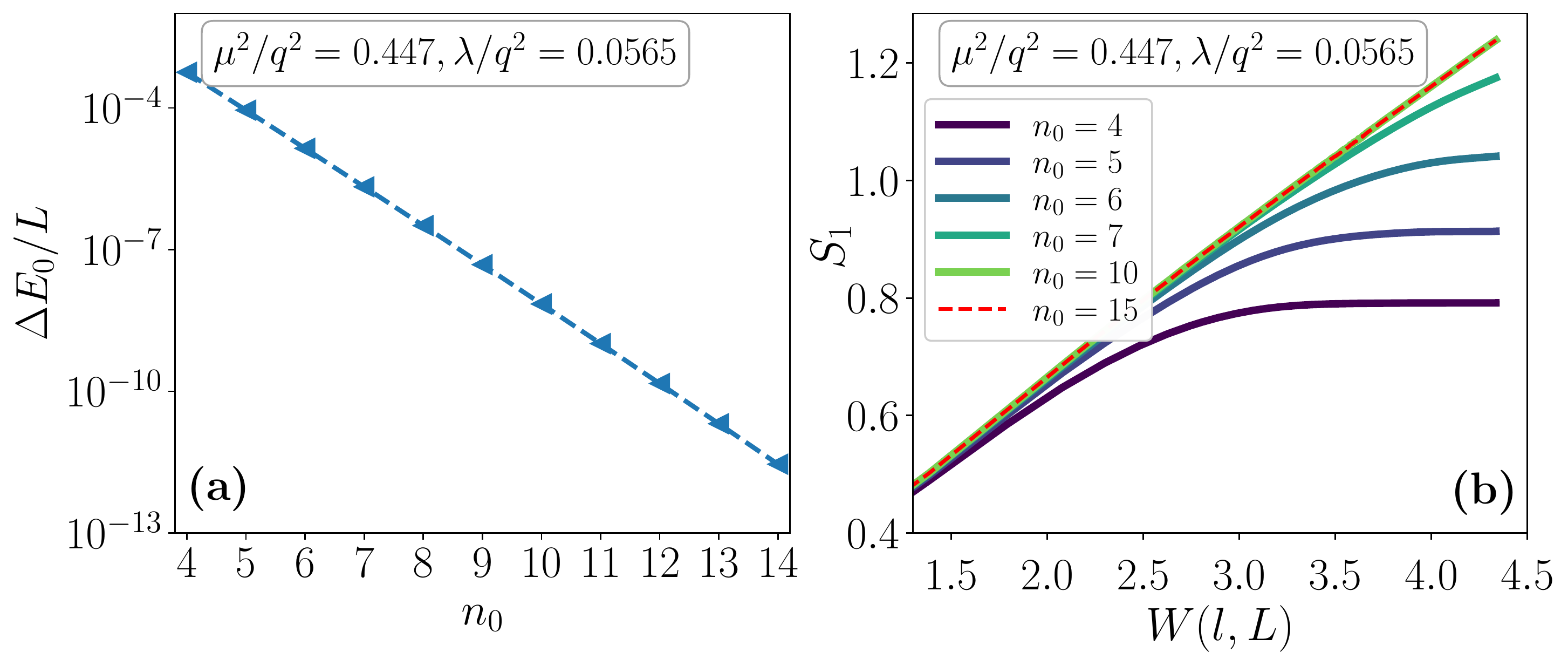}
\caption{(Color online.) \textbf{(a)} The convergence of ground-state energy density  with respect to the maximum bosonic occupancy $n_0$ for a system of size $L=120$ at the critical point. Here we plot the relative errors in the energy density $\Delta E_0/ L$, where $\Delta E_0 = |E_0({n_0}) - E_0({n_0 +1})|$.
\textbf{(b)} The profile of the von Neumann entanglement entropy $S = S_1$ with respect to the cord length $W = \ln \left(\frac{2L}{\pi} \sin(\pi l/L) \right)$ at the critical point for $L \in [40, 120]$ and for different values of $n_0$.
\label{fig:n0_convergence}
}
\end{figure}

In our MPS representation, we truncate the Hilbert spaces of both the  bosonic species, `$a$' and  `$b$', upto the maximum occupancy of $n_0 = 10$ (i.e., 11 levels of each type of bosonic matter), that results into the physical dimension of 121 for the MPS tensors on the matter sites. In order to verify the convergence with respect to this cutoff, we vary the maximum bosonic occupancy $n_0$ in the range $[4, 15]$, and check for convergence of different observables (see Fig.~\ref{fig:n0_convergence}). Specifically, we show in Fig.~\ref{fig:n0_convergence}\textbf{(a)} that for $n_0=10$ the error in the ground-state energy density for system-size $L=120$ falls below $10^{-7}$ at the critical point. Moreover, the entropy profile for $n_0=10$ is essentially identical to the one with $n_0=15$, capturing the proper entropy scaling with respect to the cord length (Fig.~\ref{fig:n0_convergence}\textbf{(b)}). For simulating short-range matter-gauge Hamiltonian~\eqref{eq:Hamil}, we truncate the gauge field Hilbert spaces to 15 levels electric basis (see Appendix~\ref{app:comparison}).

\begin{figure}
\centering
\includegraphics[width=\linewidth]{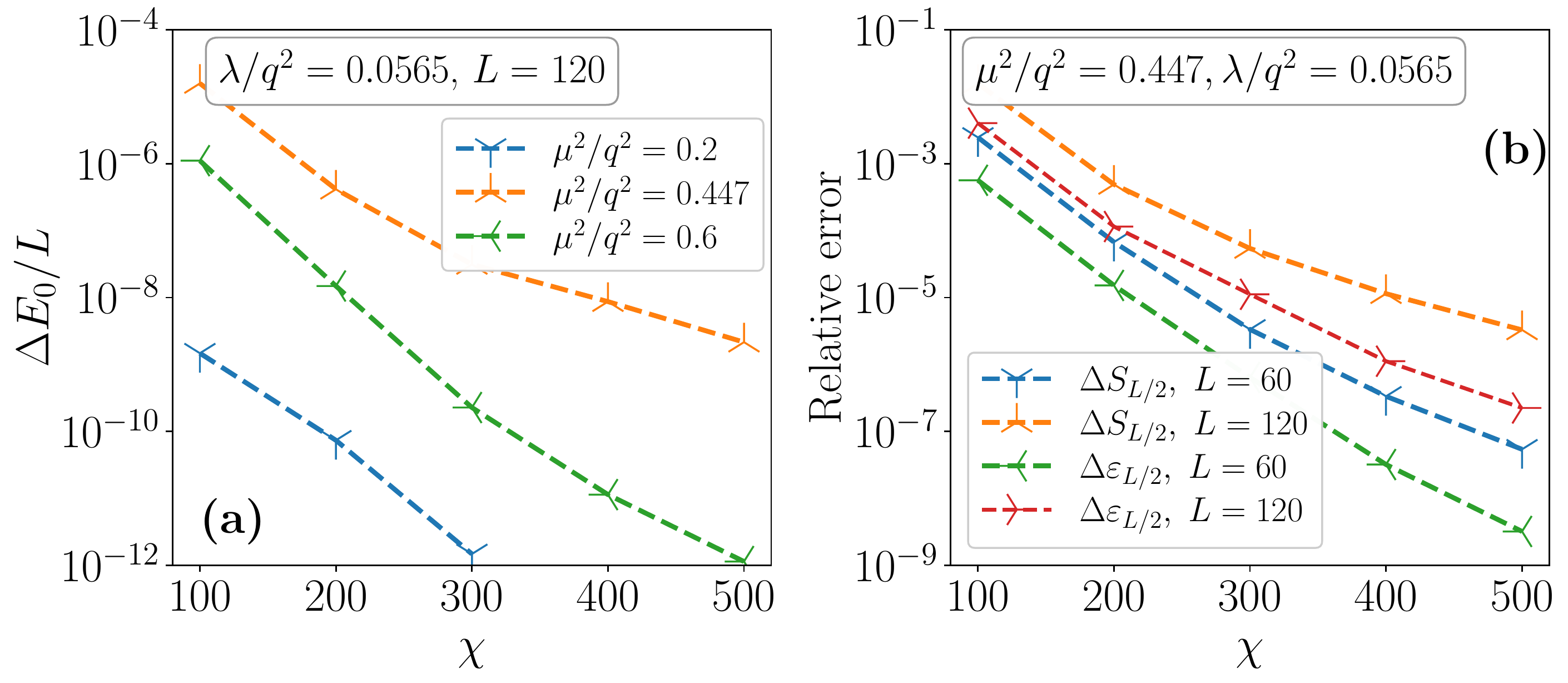}
\caption{(Color online.) The convergence of \textbf{(a)} the energy density ($\Delta E_0/L$) and \textbf{(b)} the half-chain von Neumann entanglement entropy ($\Delta S_{L/2}$) and the half-chain entanglement ground state energy ($\Delta \varepsilon_{L/2}$)  with respect to the maximum bond dimension used in the simulation, respectively $\chi \in \{100, 200, 300, 400, 500, 600\}$. Here we plot relative errors in the quantities as $\Delta \mathcal{O} = |\mathcal{O}_{\chi} - \mathcal{O}_{\chi+100}|$. In \textbf{(a)}  we choose three points in the phase diagram, namely (1) a point in the confined phase ($\mu^2/q^2 = 0.2$), (2)  the critical point ($\mu^2/q^2 = 0.447$), and (3) a point in the Higgs phase ($\mu^2/q^2 = 0.6$) for the system-size $L=120$. In \textbf{(b)} we show the convergence of the entropic quantities as a function of the bond dimension at the critical point.
\label{fig:chi_convergence}
}
\end{figure}

For the DMRG simulations, the MPS bond dimension is truncated upto $\chi=600$.  Fig.~\ref{fig:chi_convergence}\textbf{(a)} shows the convergence of ground-state energy density with respect to the bond dimension $\chi \leq 600$ for systems of sizes $L \leq 120$.
The energy density converges close to the machine precision within $\chi \leq 500$ in the gapped regions -- confined ($\mu^2/q^2 = 0.2$) and Higgs ($\mu^2/q^2$). On the other hand, as expected, the convergence is slower at the critical point due to the diverging correlation length. However, 
as shown in Fig.~\ref{fig:chi_convergence}\textbf{(b)}, the precision attained at the critical point for $\chi = 500, 600$ is sufficient to perform the precise scaling analysis reported here. To confirm the convergence of the DMRG sweeps, we continue the DMRG iterations until the energy difference in subsequent sweeps falls below $10^{-13}$.

In case of out-of-equilibrium simulations with TDVP, we truncate the bosonic species upto the maximum occupancy of $n_0=7$ (i.e, 8 levels) levels and restrict the maximum MPS bond upto $\chi = 400$.

\section{Numerical comparisons between the local matter-gauge and the matter-only long-range descriptions of the system}
\label{app:comparison}

\begin{figure}[h]
    \centering
    \includegraphics[width=0.95\linewidth]{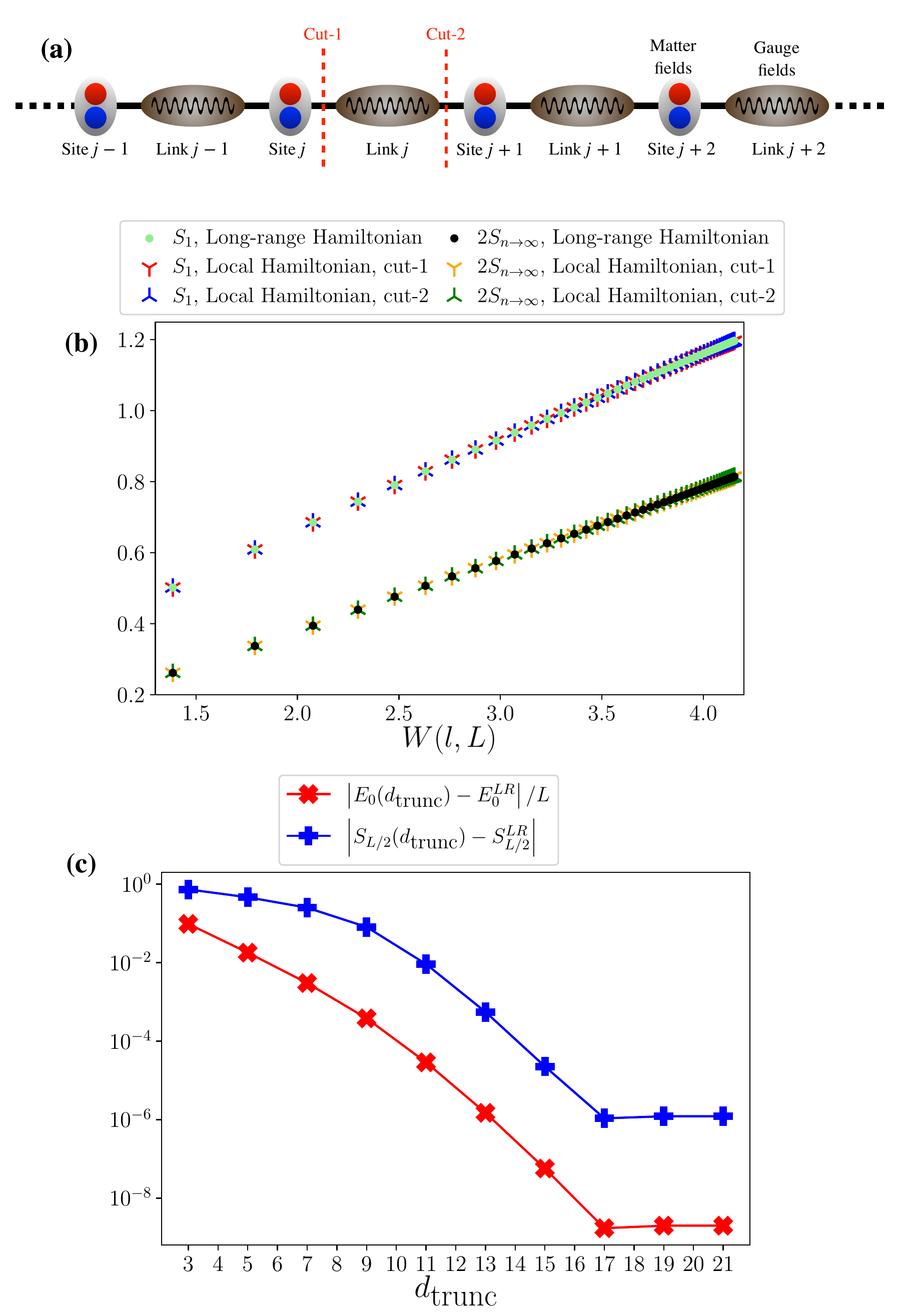}
    \caption{
\textbf{(a)} In presence of gauge fields at the links between two nearest-neighbor sites, there are two options to take bipartition at a given link, i.e., cut-1 and cut-2 depicted in the figure, for the calculation of entanglement entropies. The entropies, however, remain same between these two options since gauge fields in 1+1D are not dynamical variables.
\textbf{(b)} The comparison between von Neumann entropy $S = S_1$ and the entanglement ground state energy $\varepsilon_0 = S_{n \rightarrow \infty}$ calculated from the long-range Hamiltonian after integrating-out the gauge fields and from the original local matter-gauge Hamiltonian for system-size $L=100$ at the critical point. For the local gauge theory Hamiltonian, we truncate the gauge-field Hilbert spaces upto $d_{\text{truc}} = 21$ electric states.
\textbf{(c)} We plot the absolute differences in ground state energy densities and in the mid-chain von Neumann entanglement entropies between the truncated local matter-gauge Hamiltonian with respect to the long-range system as a function of the truncated gauge-field Hilbert-space dimension $d_{\text{trunc}}$. Here we consider the system-size $L=60$.
The differences reduce to the numerical accuracy attained with the MPS bond dimension $\chi=600$ at the truncation level $d_{\text{trunc}} = 17$.
}
\label{fig:ent_comp}
\end{figure}

In the main text, we have favored the long-range Hamiltonian \eqref{eq:Hamil_pureM} after integrating-out the gauge fields by an exact mapping utilizing the Gauss law. This is because of the advantage in the long-range formulation, where we avoid dealing with extra degrees of freedom corresponding to the gauge fields. Moreover, since the Hilbert spaces of the gauge fields are formally infinite dimensional, they need to be truncated to a finite dimensional subspace in the short-range formulation of Eq.~\eqref{eq:Hamil}, thereby introducing an extra source of approximation that needs to be accurately controlled (see e.g.,~\cite{kuhn_pra_2014}).
Here, we show that the short-range Hamiltonian, with sufficiently large truncated gauge-field Hilbert spaces, gives the exactly same results as that of the long-range Hamiltonian discussed in the main text.

To be noted that, in presence of the gauge fields in the short-range formulation, we have two options to take bipartitions at each link -- one before the gauge field (cut-1) and another after the gauge field (cut-2) -- for the calculation of entanglement entropies as seen in Fig.~\ref{fig:ent_comp}\textbf{(a)}. 
In Figs.~\ref{fig:ent_comp}\textbf{(b)}-\textbf{(c)}, we compare the results between the truncated short-range matter-gauge Hamiltonian \eqref{eq:Hamil} and the long-range Hamiltonian \eqref{eq:Hamil_pureM}. First, we can observe that in case of local gauge theory, entropies calculated at cut-1 or cut-2 are the same, which is
expected since the gauge fields are not dynamical variables in 1+1D, see Fig.~\ref{fig:ent_comp}\textbf{(b)}. 
Moreover, with sufficiently large gauge-field Hilbert space dimension ($d_{\text{trunc}}$), local and long-range Hamiltonians give exactly same result (upto a numerical accuracy, see attached Fig.~\ref{fig:ent_comp}\textbf{(c)}).
This is understandable as in 1+1D, integrating-out the gauge fields using Gauss law is exact.
Specifically, Fig.~\ref{fig:ent_comp}\textbf{(c)} shows that the differences between the results coming from truncated local gauge theory Hamiltonian and the long-range Hamiltonian diminute to the numerical accuracies (attained with the MPS bond dimension $\chi=600$) for $d_{\text{trunc}} \geq 17$.

\acknowledgments

We acknowledge the PL-GRID infrastructure for providing us the high-performance computing facility under the grant PLG/2022/015613 for 
the numerical results reported here. The MPS algorithms have been implemented using ITensor library~\cite{itensor, itensor-r0.3}.

The work of M.D. was partly supported by the ERC under grant number 758329 (AGEnTh), by the MIUR Programme FARE (MEPH), and by the European Union's Horizon 2020 research and innovation programme under grant agreement No 817482 (Pasquans).

M.L. acknowledges support from: ERC AdG NOQIA; Ministerio de Ciencia y Innovation Agencia Estatal de Investigaciones (PGC2018-097027-B-I00/10.13039/501100011033, CEX2019-000910-S/10.13039/501100011033, Plan National FIDEUA PID2019-106901GB-I00, FPI, QUANTERA MAQS PCI2019-111828-2, QUANTERA DYNAMITE PCI2022-132919, Proyectos de I+D+I “Retos Colaboraci\'on” QUSPIN RTC2019-007196-7); MICIIN with funding from European Union NextGenerationEU(PRTR-C17.I1) and by Generalitat de Catalunya; Fundació Cellex; Fundaci\'o Mir-Puig; Generalitat de Catalunya (European Social Fund FEDER and CERCA program, AGAUR Grant No. 2021 SGR 01452, QuantumCAT \ U16-011424, co-funded by ERDF Operational Program of Catalonia 2014-2020); Barcelona Supercomputing Center MareNostrum (FI-2022-1-0042); EU Horizon 2020 FET-OPEN OPTOlogic (Grant No 899794); EU Horizon Europe Program (Grant Agreement 101080086 — NeQST), ICFO Internal “QuantumGaudi” project; European Union’s Horizon 2020 research and innovation program under the Marie-Skłodowska-Curie grant agreement No 101029393 (STREDCH) and No 847648 (“La Caixa” Junior Leaders fellowships ID100010434: LCF/BQ/PI19/11690013, LCF/BQ/PI20/11760031, LCF/BQ/PR20/11770012, LCF/BQ/PR21/11840013).

This research was also funded by National Science Centre (Poland)   by grant 2021/03/Y/ST2/00186 within the QuantEra II Programme that has received funding from the European Union's Horizon 2020 research and innovation programme under Grant Agreement No 101017733 DYNAMITE  (J.Z.). The partial support by the Strategic Programme Excellence Initiative at Jagiellonian University is also acknowledged. 

LT acknowledges the support from the Spanish project
PGC2018-095862-B-C22, the CSIC Interdisciplinary Thematic Platform (PTI) Quantum Technologies (PTI-QTEP+) and from the Spanish projects PID2021-127968NB-I00 and TED2021-130552B-C22 funded by r MCIN/AEI/10.13039/501100011033/FEDER, UE and MCIN/AEI/10.13039/501100011033, respectively.

Views and opinions expressed in this work are, however, those of the authors only and do not necessarily reflect those of the European Union, European Climate, Infrastructure and Environment Executive Agency (CINEA), nor any other granting authority. Neither the European Union nor any granting authority can be held responsible for them.

\bibliographystyle{apsrev4-1}
\bibliography{revised_prb.bbl}

\end{document}